\documentclass[a4paper,11pt]{article}

\usepackage{jheppub,bm} 

\usepackage[T1]{fontenc} 
\usepackage{slashed}

\usepackage{graphicx}
\def\Maryland{Maryland Center for Fundamental Physics, University of Maryland, College Park, Maryland 20742, USA}
\def\Argonne{High Energy Physics Division, Argonne National Laboratory, Argonne, IL 60439, USA}
\def\Northwestern{Department of Physics \& Astronomy, Northwestern University, Evanston, IL 60208, USA}
\def\Fermilab{Fermilab, P.O.Box 500, Batavia, IL 60510, USA}
\def\Durham{Institute for Particle Physics Phenomenology, Department of Physics, University of Durham, Durham, DH1 3LE, UK}
\def\Buffalo{Department of Physics, University at Buffalo, The State University of New York, Buffalo 14260 USA}
\preprint{ FERMILAB-PUB-16-120-T, IPPP/16/32}
\title{Color singlet production at NNLO in MCFM}

\author{Radja Boughezal$^{a}$, }    
\author{John M. Campbell$^{b}$, }
\author{R. Keith Ellis$^{c}$, }
\author{Christfried Focke$^{d}$,}
\author{Walter Giele$^{b}$, }     
\author{Xiaohui Liu$^{e}$, }
\author{Frank Petriello$^{a,d}$}     
\author{and Ciaran Williams$^{f}$}

\affiliation[a]{\Argonne}
\affiliation[b]{\Fermilab}
\affiliation[c]{\Durham}
\affiliation[d]{\Northwestern}
\affiliation[e]{\Maryland}
\affiliation[f]{\Buffalo}

\emailAdd{rboughezal@anl.gov}
\emailAdd{johnmc@fnal.gov}
\emailAdd{keith.ellis@durham.ac.uk}
\emailAdd{christfried.focke@northwestern.edu}
\emailAdd{giele@fnal.gov}
\emailAdd{xhliu@umd.edu}
\emailAdd{f-petriello@northwestern.edu}
\emailAdd{ciaranwi@buffalo.edu}

\def\psing{p_{{\rm singlet}}}
\def\taun{{\cal T}_N}
\def\tauzero{{\cal T}_0}
\def\tauncut{{\cal T}_N^{cut}}
\def\tauzerocut{{\cal T}_0^{cut}}
\newcommand{\rd}{{\rm d}}

\newcommand{\beq}{\begin{equation}}
\newcommand{\eeq}{\end{equation}}

\abstract{ 
We present the implementation of several color-singlet final-state processes at
Next-to-Next-to Leading Order (NNLO) accuracy in QCD to the publicly available parton-level Monte Carlo program MCFM\footnote{Version 8.0 of MCFM can be
downloaded from the {\tt mcfm.fnal.gov} website.}. 
Specifically we discuss the processes $pp\rightarrow H$,
$pp\rightarrow Z$, $pp\rightarrow W$, $pp\rightarrow HZ$,
$pp\rightarrow HW$ and $pp\rightarrow\gamma\gamma$. Decays of
the unstable bosons are fully included, resulting in a flexible fully differential Monte Carlo code. 
The NNLO corrections have been calculated
using the non-local $N$-jettiness subtraction approach. 
Special attention is given to the numerical
aspects of running MCFM for these processes at this
order. We pay particular attention to the systematic uncertainties due
to the power corrections induced by the $N$-jettiness regularization scheme and the evaluation time needed to run the hybrid
openMP/MPI version of MCFM at NNLO on multi-processor systems.}

\begin{document} 
\maketitle
\flushbottom

\section{Introduction}

The second run of the LHC (Run II) which is currently underway, will
result in the accumulation of an unprecedented amount of high-quality
data in a new high energy regime.  In tandem with the well-understood
and carefully calibrated detectors, this will lead to experimental
uncertainties that are at the level of a few percent or smaller for
many of the most important processes.  These include various Higgs
boson production channels, as well as standard candle processes such
as vector boson production. Studies of diboson production will allow
for stringent tests of the Electroweak sector of the Standard Model (SM) and
constraints on possible new physics scenarios.  In order to make best
use of the precise experimental observations it is crucial to have
access to accurate theoretical calculations of the same quantities.
At the LHC this requires the calculation of QCD corrections to
inclusive and differential cross sections at increasingly higher
order. For the most efficient comparison between theoretical
predictions and experimental data it is extremely beneficial for
theoretical results to be released in the form of a public code,
allowing users full flexibility in obtaining theoretical predictions
relevant for their analysis.

While calculations at Next-to-Leading Order (NLO) in the strong
coupling constant are by now quite standard, only about 20
processes have been calculated through to Next-to-Next-to-Leading
Order (NNLO). Recent publications on these processes are shown in Table.~\ref{NLLOcites}.
\begin{table}[b]
\begin{center}
\begin{tabular}{|l|l|l|l|}
\hline
$H+ 0$~jet~\cite{Anastasiou:2004xq,Anastasiou:2005qj,Catani:2007vq,Gaunt:2015pea}
&  $H+ 1$~jet~\cite{Boughezal:2013uia,Chen:2014gva,Boughezal:2015dra,Boughezal:2015aha,Chen:2016vqn}
& Higgs WBF~\cite{Cacciari:2015jma}  
& $H \to b \bar{b}$~\cite{Anastasiou:2011qx,DelDuca:2015zqa} \\
$W+ 0$~jet~~\cite{Melnikov:2006di,Catani:2009sm}
& $Z/\gamma^* + 0$~jet~~\cite{Melnikov:2006kv,Catani:2009sm,Gaunt:2015pea}
& $W+ 1$~jet~\cite{Boughezal:2015dva}  
& $Z+ 1$~jet~\cite{Ridder:2015dxa,Boughezal:2015ded,Ridder:2016rzm,Ridder:2016nkl} \\
$ZH$~\cite{Ferrera:2014lca,Campbell:2016jau} 
& $WH$~\cite{Ferrera:2013yga,Campbell:2016jau}  
& $WZ$~\cite{Grazzini:2016swo}
& \\
$ZZ$~\cite{Cascioli:2014yka,Grazzini:2015hta,Caola:2015psa} 
& $WW$~\cite{Gehrmann:2014fva,Caola:2015rqy,Grazzini:2016ctr}  
&$W+\gamma,Z+\gamma$~\cite{Grazzini:2016hai}  
& $\gamma \gamma$~\cite{Catani:2011qz,Campbell:2016yrh} \\
$t \bar{t}$~\cite{Czakon:2016ckf,Abelof:2015lna}  
& single top~\cite{Brucherseifer:2014ama}  
& top decay~\cite{Gao:2012ja,Brucherseifer:2013iv}  
& dijets~\cite{Currie:2014upa}  \\
\hline
\end{tabular}
\end{center}
\caption{Publications on processes evaluated differentially at NNLO.}
\label{NLLOcites} 
\end{table}
 
All such calculations require a means by
which to regulate the soft and collinear radiation that appears in the
calculation of the higher-order contributions.  At NLO local
subtraction schemes, such as FKS~\cite{Frixione:1995ms} or
Catani-Seymour dipole subtraction~\cite{Catani:1996vz}, are typically
preferred.  In these local subtraction formalisms, the singular
unresolved infra-red limits are cancelled point-wise by local
counterterms. These local counterterms, after analytic integration
over the unresolved partons, are added to the virtual
corrections yielding a finite result.

The construction of a local subtraction scheme for a NNLO calculation
is a daunting task, given the complexities of the multiple infrared limits
and differing dimensionality of phase space for the component
parts. However, progress has been made, with significant advances over
the last decade.  The first local subtraction scheme used at NNLO was
the sector decomposition approach presented in
ref.~\cite{Anastasiou:2003gr}. This scheme separates the overlapping
singularities by using a plus-prescription to isolate the singular
contributions, thereby avoiding any analytic integrations over regions
of phase space.  The antenna subtraction method was extended to NNLO
in Refs.~\cite{Somogyi:2005xz,GehrmannDeRidder:2005cm}, and has been
used to obtain predictions for $2\rightarrow 2$ processes in which
both final state particles are colored~\cite{Currie:2014upa}.  Antenna
subtraction resembles the NLO subtraction formalisms in that the
doubly unresolved limits are cancelled point-by-point in phase space
by counterterms which require analytic integration to cancel infrared poles
in the real-virtual and double virtual phase spaces.  Finally in
Refs.~\cite{Czakon:2010td,Czakon:2011ve,Boughezal:2011jf} the sector decomposition
approach was generalized to arbitrary processes. By partitioning the
phase space into appropriate sectors in which each singularity can be made manifest, 
and then performing a Laurent series expansion to extract
the poles. This method has been applied to various processes at the
LHC~\cite{Brucherseifer:2013iv,Boughezal:2013uia,Boughezal:2015dra,Czakon:2016ckf}.

In addition to the local subtraction schemes discussed above, there is
an alternate form of regulation, which is inherently non-local.
Indeed one of the first NLO regularization techniques developed was
one such method, phase space slicing, introduced in
Refs.~\cite{Baer:1989jg,Giele:1991vf,Giele:1993dj}.  In these methods
a parameter is used to separate the resolved and unresolved phase
spaces. The resolved region of phase space corresponds to a
calculation of the process with one additional final state parton, and
if a suitable resolution parameter is chosen, the unresolved region
can be directly calculated.  At NLO non-local methods have generally
fallen out of favor.  This is due to the large cancellation
between the resolved and unresolved contributions at small values of
the resolution parameter, which can induce large Monte Carlo
uncertainties.  

However non-local subtraction schemes have
made a resurgence for NNLO calculations.  Although they have the
disadvantages discussed above they also have several advantages which
make them attractive for NNLO calculations.  First, they are
conceptually simple to implement. Once a suitable resolution parameter
is selected, the singly unresolved part of the calculation can be
obtained with existing NLO event generators, such as
MCFM~\cite{Campbell:1999ah,Campbell:2011bn,Campbell:2015qma}.  Second, with recent advances in
computing, the drawback associated with the large numerical
cancellations can be mitigated by running with a large number of
computer cores.  Finally by using a resolution parameter motivated by a 
physical factorization theorem, the approximations inherent in the method can be systematically improved, e.g. by analytic 
calculations of power-suppressed contributions~\cite{BPL}.


The first non-local subtraction developed for NNLO calculation was the
so-called $q_T$ subtraction method~\cite{Catani:2007vq}.  This method
uses the transverse momentum of the final state color neutral
particle, $q_T$, as the cut variable. For $q_T < q_T^{\rm{cut}}$ the
factorization theorem of Collins, Soper and
Sterman~\cite{Collins:1984kg}, can be used to compute the cross
section, while above the cutoff the NLO calculation of the
color-singlet plus jet can be utilized.
An obvious drawback is that it is only applicable to color neutral
final states.  
 Inspired by a factorization formula~\cite{Beneke:2002ph} from Soft Collinear Effective Field Theory
(SCET)~\cite{Bauer:2000ew,Bauer:2000yr,Bauer:2001ct,Bauer:2001yt,Bauer:2002nz} the first steps towards extending these ideas to
calculations containing colored final states were taken in a
calculation of top-quark decay at NNLO~\cite{Gao:2012ja}.  
However, no initial collinear singularities appear in this calculation.  A powerful generalization of this idea applicable to general initial and final states was introduced in~\cite{Boughezal:2015dva,Gaunt:2015pea}.  It is obtained by replacing the $q_T$
variable with the event shape $N$-jettiness
variable~\cite{Stewart:2010tn}.  Below the $N$-jettiness ($\tau_N$)
cutoff, SCET
provides the relevant factorization theorem~\cite{Stewart:2010tn}.  For the below-cut region
the necessary SCET ingredients, corresponding to the final state and
initial state collinear radiation functions are already known, and are
represented by the two-loop jet-functions of
Ref.~\cite{Becher:2006qw,Becher:2010pd} and the two-loop beam-function
of Ref.~\cite{Gaunt:2014xga,Gaunt:2014cfa}.  The corresponding
two-loop soft functions are also known for
zero-jettiness~\cite{Kelley:2011ng,Monni:2011gb} and for general 
$N$-jettiness~\cite{Boughezal:2015eha}.  The first process calculated
at NNLO using this method was $pp\rightarrow
W$+jet~\cite{Boughezal:2015dva},  followed by calculations of
the $pp\rightarrow$ Higgs+jet~\cite{Boughezal:2015aha} and
$pp\rightarrow Z$+jet~\cite{Boughezal:2015ded} processes, and by detailed phenomenological 
studies of these processes at this order~\cite{Boughezal:2016yfp,Boughezal:2016dtm,Boughezal:2016isb}.  
$pp\rightarrow H$ and $pp\rightarrow Z$ were also calculated using this
method~\cite{Gaunt:2015pea}. Additional processes of phenomenological interest, $pp\rightarrow
VH$~\cite{Campbell:2016jau} and $pp\rightarrow
\gamma\gamma$~\cite{Campbell:2016yrh} have been calculated using the same approach. 

As mentioned before, an important advantage of the $N$-jettiness
subtraction method is that it meshes well with the existing NLO
calculations, such as those included in MCFM.
Included in MCFM are the NLO corrections to $W+n~$ jet production,
$Z+n~$ jets production~\cite{Campbell:2002tg}, Higgs~$+n$ jets
production~\cite{Campbell:2006xx,Campbell:2010cz} for ($n=0,1,2$),
making the implementation of $W$, $Z$, $H$ + 0,1 jet at NNLO possible.

The recent advances in NNLO technologies  allows for the exciting possibility 
of releasing a public code capable of computing many $2\rightarrow 2$ processes at NNLO accuracy. 
This paper presents a first step in this journey by summarizing the implementation of the $N$-jettiness subtraction procedure
in MCFM, and presenting a detailed breakdown of the method for the processes released in the initial version
of the NNLO code.  An important consideration in making the code public is computational
speed. In Ref.~\cite{Campbell:2015qma}  MCFM  was  upgraded to use a parallel
version of the VEGAS adaptive integration method using openMP. 
For NNLO calculations, this was
expanded by using a hybrid openMP/MPI version of MCFM for use on computing clusters
to facilitate the numerical NNLO calculations of
Ref.~\cite{Boughezal:2015ded}. Using the hybrid version of MCFM we
can calculate NNLO distributions efficiently within a reasonable timescale. 

In summary, this paper describes the implementation of the $N$-jettiness subtraction procedure in MCFM and presents 
results for the processes available in MCFM v8.0. Specifically these processes are $pp\rightarrow H,W,Z,VH,\gamma\gamma$. 
Where present the decays of unstable particles are included, allowing for a fully flexible MC code. 
In section~\ref{sec:SCET} we will give a schematic overview of the non-local $N$-jettiness subtraction scheme.
Section~\ref{sec:overview} will detail the calculational set-up and sections~\ref{sec:nlo} and~\ref{sec:nnlo} will look at the 
$N$-jettiness subtraction at NLO and NNLO respectively. A preliminary study of the effects of the dominant power corrections is presented in Section~\ref{power}.
The more numerical aspects are studied
in section~\ref{sec:comp}. Finally in section~\ref{sec:conc} the main results are summarized.
\section{SCET Based Non-Local Subtraction}
\label{sec:SCET}
A collision of partons $a$ and $b$ with momentum fractions $x_{a,b}$,
originating from the incoming beam protons with momenta  $p_{a,b}$, 
produces a final state including $N$ jets with momenta $\{q_i\}$. The
jettiness of parton $j$ with momentum $p_j$ is defined as
\beq
\taun(p_j)=\min_{i=a,b,1,\ldots,N}\left\{\frac{2\,q_i\cdot p_j}{{Q_i}}\right\}\ ,
\eeq
where for notational simplicity we have set $q_{a,b}=p_{a,b}$.   
We denote the jet or beam energy by $E_i$.
$Q_i$ is a measure of the jet/beam hardness.  
In our numerical results we set this equal to twice the jet/beam energy, 
$Q_i=2 E_i$~\cite{Stewart:2010tn}.
We can now define the event jettiness, or $N$-jettiness, as the sum
over all the $M$ final state parton jettiness values
\beq
\taun=\sum_{k=1}^M\taun(p_k)=\sum_{k=1}^M
\min_{i=a,b,1,\ldots,N}\left\{\frac{2\,q_i\cdot p_k}{ { Q_i}}\right\}\ . 
\eeq
For Leading Order (LO) events we have $\{p_i\}=\{q_i\}$ and the event jettiness is zero. 
Beyond LO ($M>N$), only in the
soft/collinear limit will the event jettiness  necessarily go to zero. Therefore the
event $N$-jettiness can be used in a non-local subtraction approach where we can isolate
the doubly unresolved region of the phase space by demanding $\taun<\tauncut$.
In this paper we restrict ourselves to color singlet final state events. We can therefore use the
event shape variable $\tauzero$ to regulate the  initial state radiation. 

By demanding $\tauzero<\tauzerocut$ one isolates the doubly unresolved
regions of phase space.  The matrix elements in the soft/collinear
approximation can be analytically integrated over this region and
added to the virtual contributions.  The regions of phase space where
$\tauzero>\tauzerocut$ are integrated over numerically.  In the limit
$\tauzerocut\rightarrow 0$ this will result in the correct 
results for the cross section.

To obtain the analytic soft/collinear expressions we use all-orders resummation results 
which rely heavily on the machinery of soft-collinear 
effective theory (SCET)~\cite{Bauer:2000ew,Bauer:2000yr,Bauer:2001ct,Bauer:2001yt,Bauer:2002nz}.  
The all-orders resummation of the $\tauzero$ event-shape variable in the 
limit $\tauzero \rightarrow 0$ was constructed in Ref.~\cite{Stewart:2010tn}:
\begin{equation}
\label{eq:SCETfac1}
\frac{\rd \sigma}{\rd \tauzero} =\sum_{ab} \int \rd x_a \rd x_b \int \rd
\Phi_B(p_a,p_b;\psing)\,\Theta(\psing)\,  H_{ab}(\Phi_B,\mu)\,\frac{\rd\Delta_{ab}}{\rd\tauzero}
+\ldots\ ,
\end{equation}
where the indices $a,b$ run over all initial state partons involved in the
scattering. The initial state momenta $p_{a,b}$ are given by the momenta fractions $x_{a,b}$, while 
$\Phi_B$ denotes the Born-level color singlet phase space $p_ap_b\rightarrow\psing$.  
The composite $\Theta(\psing)$ denotes any phase space restrictions on the color-singlet phase space.   
The soft/collinear function $\Delta_{ab}$ is given by
\begin{eqnarray}
\label{eq:SCETfac2}
\frac{\rd\Delta_{ab}}{\rd\tauzero}&=&B_a\otimes B_b\otimes S_{ab}\nonumber\\
&\equiv& \int \rd t_{B_a} \rd t_{B_b}  \rd t_S \, \delta\left(
 \tauzero-t_{B_a}-t_{B_b}-t_S\right) 
 \, B_{a}(t_{B_a},x_a,\mu) \, B_{b}(t_{B_b},x_b,\mu)
         \,S_{ab}(t_S,\mu)\ . \nonumber\\
\end{eqnarray}  
A summary of the various components which appear in these expressions  is given below:
\begin{itemize}

\item The hard function $H$ encodes the effect of hard virtual corrections.  At leading order in the
$\alpha_s$-expansion it reduces to the leading-order partonic cross section.  At higher orders it also contains
the finite contributions of the pure virtual corrections, renormalized
at scale $\mu$ using the $\overline{\text{MS}}$ scheme. 
It depends on the Born-level kinematics and the scale choice.

\item The beam function $B_a$ contains the effects of initial-state collinear radiation.  It depends on
$t_{B_a}$, the contribution of initial-state collinear radiation to $\tauzero$.  The beam function is
non-perturbative; however, up to corrections suppressed by $\Lambda_{\text{QCD}}/t_B$, it can be written as a
convolution of perturbative matching coefficients and the usual parton
density functions, $f_{i/H}$,
\begin{equation} B_a(t_{B_a},x,\mu) = \sum_i
\int_x^1 \frac{\rd \xi}{\xi} {\cal I}_{ai} (t_{B_a},x/\xi,\mu) f_{i/H}(\xi),
\end{equation}
where we have suppressed the scale dependence of the parton density
functions, and $i$ runs over all partons.  The two-loop beam functions
have been computed in Refs.~\cite{Gaunt:2014xga,Gaunt:2014cfa}.

\item The soft function $S$ collects the jettiness contributions of soft radiation.   It depends on $t_S$, the
contribution of soft radiation to $\tauzero$.
The expansion of the soft function for zero-jettiness up to two-loop order can be found in Refs.~\cite{Kelley:2011ng,Monni:2011gb}.
\end{itemize}
The delta function appearing in Eq.~(\ref{eq:SCETfac2}) combines the
contribution of each type of radiation to produce the measured value
of $\tauzero$.  The factorization formula is correct up to power
corrections, indicated by the ellipsis in Eq.~(\ref{eq:SCETfac1}). 
These power corrections can in principle be calculated in the same way as one derives the leading power components in Eq.~(\ref{eq:SCETfac2}). However, they can be neglected as long as we restrict ourselves to the
phase-space region $\tauzero \ll Q$, where $Q$ denotes the hard scale
in the process (for the zero-jet processes considered here, $Q$ is of
the order of the invariant mass of the final state).  Integrating
Eq.~(\ref{eq:SCETfac1}) over the region $\tauzero <\tauzerocut$ will
give the analytic result for the below-cut cross section:
\begin{equation}
\rd\sigma(\tauzerocut) =\sum_{ab} \int \rd x_a \rd x_b \int \rd
\Phi_B(p_a,p_b;\psing)\,\Theta(\psing)\,  H_{ab}(\Phi_B,\mu)\,\Delta_{ab}(\tauzerocut)
+\cdots\ ,
\end{equation}
with
\begin{equation}
\Delta_{ab}(\tauzero<\tauzerocut)=\int_0^{\tauzerocut}\rd\tauzero\
\left( B_a\otimes B_b\otimes S_{ab}\right)\ .
\end{equation}

Next we expand the functions order by order in $\alpha_S$ using a
superscript to denote the power of $\alpha_s$ appearing in each term.
That is, we expand any perturbative function ${\cal F}$ as
\begin{equation}
{\cal F} = {\cal F}^{(0)}+\alpha_S{\cal F}^{(1)}+\alpha_S^2{\cal
  F}^{(2)}+\cdots\, .
\end{equation}
This results in
\begin{eqnarray}
\rd\sigma^{(n)}(\tauzerocut) &=&\sum_{ab} \int \rd x_a \rd x_b \int \rd
\Phi_B(p_a,p_b;\psing)\,\Theta(\psing)\sum_{k=0}^n  H_{ab}^{(n-k)}(\Phi_B,\mu) \Delta_{ab}^{(k)}(\tauzerocut)\ ,\nonumber\\
\Delta_{ab}^{(n)}(\tauzerocut)&=&\sum_{k+l+m=n}\int_0^{\tauzerocut}\rd\tauzero\left( B_a^{(k)}\otimes B_b^{(l)}\otimes S_{ab}^{(m)}\right)\ .
\end{eqnarray}

To obtain the ${\cal O}(\alpha_s^2)$ correction to the soft/collinear cross section below the $\tauzerocut$
we need 
\begin{eqnarray}
\Delta_{ab}^{(0)}(\tauzerocut)&=&\int_0^{\tauzerocut}\rd\tauzero\ B_a^{(0)}\otimes B_b^{(0)}\otimes S_{ab}^{(0)}=f_{a/H}(x_a)\,f_{b/H}(x_b)\ ,\nonumber\\
\Delta_{ab}^{(1)}(\tauzerocut)&=&\int_0^{\tauzerocut}\rd\tauzero\left(
 B_a^{(1)}\otimes B_b^{(0)}\otimes S_{ab}^{(0)}+ B_a^{(0)}\otimes B_b^{(1)}\otimes S_{ab}^{(0)}+ B_a^{(0)}\otimes B_b^{(0)}\otimes S_{ab}^{(1)}
\right)\ ,\nonumber\\
\Delta_{ab}^{(2)}(\tauzerocut)&=&\int_0^{\tauzerocut}\rd\tauzero\left(
 B_a^{(2)}\otimes B_b^{(0)}\otimes S_{ab}^{(0)}+ B_a^{(0)}\otimes B_b^{(2)}\otimes S_{ab}^{(0)}+ B_a^{(0)}\otimes B_b^{(0)}\otimes S_{ab}^{(2)}\right.\nonumber\\
&&\phantom{\int_0^{\tauzerocut}\rd\tauzero}\left. + B_a^{(1)}\otimes B_b^{(1)}\otimes S_{ab}^{(0)}+ B_a^{(1)}\otimes B_b^{(0)}\otimes S_{ab}^{(1)}+ B_a^{(0)}\otimes B_b^{(1)}\otimes S_{ab}^{(1)}
\right)\ .\nonumber\\ \label{eq:belowcutexp}
\end{eqnarray}
Note that the leading-order expressions for the subtraction functions are
proportional to delta functions in their respective hadronic variable:
\begin{equation}
B_a^{(0)}(t_{B_a},x_a,\mu)= \delta(t_{B_a}) f_{a/H}(x_a);\ S^{(0)}_{ab}(t_s,\mu)=\delta(t_s)\ .
\end{equation} 
The soft and beam function have the generic forms
\begin{eqnarray}
S^{(n)}(t_s,\mu)&=&s_{-1}^{(n)}\delta(t_s)+\sum_{k=0}^{2n-1}s_k^{(n)}\,{\cal L}_k(t_s,\mu)\ , \nonumber\\
{\cal I}_{ij}^{(n)}(t_a,z,\mu)&=&i_{-1,ij}^{(n)}(z)\delta(t_a)+\sum_{k=0}^{2n-1}i_{k,ij}^{(n)}(z)\,{\cal L}_k(t_a,\mu)\ , \nonumber\\
{\cal L}_n(t,\mu)&=&\frac{1}{\mu} \left[ \frac{\mu \,\text{ln}^n(t)}{t}\right]_+\ .
\end{eqnarray}
For example, the following contributions to the NNLO SCET function become
\begin{equation*}
\int_0^{\tauzerocut}\rd\tauzero\ S^{(2)} \otimes {\cal I}_{ai}^{(0)} \otimes {\cal I}_{bj}^{(0)} = 
\delta_{ai} \,\delta_{bj}\left\{ s_{-1}^{(2)}+\sum_{n=0}^3 \frac{1}{n+1} \,s_{n}^{(2)}\, L^{n+1} \right\}\ ,
\end{equation*}
\begin{eqnarray}
\lefteqn{\int_0^{\tauzerocut}\rd\tauzero\ S^{(1)} \otimes {\cal I}_{ai}^{(1)} \otimes {\cal I}_{bj}^{(0)} =}\nonumber\\ &&
\delta_{bj} \left\{ s_{-1}^{(1)} i^{(1)}_{-1,ai}(z)+ s_{-1}^{(1)} \, \sum_{n=0}^1 \frac{1}{n+1} \,i^{(1)}_{n,ai}(z)\, L^{{ n+1} } 
+ i^{(1)}_{-1,ai}(z) \,  \sum_{n=0}^1\frac{1}{n+1} \,s_{n}^{(1)}\, L^{ n+1} \right. \nonumber \\ && \left.
\phantom{\delta_{bj} \Big\{}+\sum_{m,n=0}^1
                                                                                                    s_m^{(1)}i^{(1)}_{n,ai}(z)\Gamma_{m,n}\right\}\ ,
\end{eqnarray} 
where
\begin{equation}
L=\text{ln}  \left( \frac{\tauzerocut }{\mu}\right)\ .
\end{equation}
and
\begin{eqnarray}
\Gamma_{0,0}&=& L^2 - \zeta_2\nonumber \\
\Gamma_{1,0}&=&\Gamma_{0,1}=\frac{L^3}{2} - \zeta_2 L + \zeta_3\nonumber \\
\Gamma_{1,1}&=& \frac{L^4}{4} -\zeta_2 L^2 + 2 \zeta_3 L -
                \frac{\zeta_2^2}{10}\ .
\end{eqnarray}
Using these results it is possible to analytically compute all of the necessary hadronic integrals in
Eq.~(\ref{eq:belowcutexp}).  The remaining integrals are over the Born phase space and parton distribution
functions, and are simple to perform numerically.   This completes the calculation of the $\tauzero < \tauzerocut$
phase space region.  We note that the cross section below $\tauzerocut$ will contain terms of the form
$\text{ln}^n (\tauzerocut)$, where $n$ ranges from 0 to 4 at NNLO.  An important check of our framework is the
cancellation of these terms against the identical logarithms that appear for $\tauzero > \tauzerocut$.  We must
also choose $\tauzerocut$ small enough to suppress the power corrections in Eq.~(\ref{eq:SCETfac1}).
 Both of these issues will be addressed in the subsequent sections.

\section{Process Overview}
\label{sec:overview}
For all of the studies performed in this paper we perform calculations for the
LHC operating at a center-of-mass energy of $\sqrt s=13$~TeV.  The parameters
that are used throughout this paper are shown in Table~\ref{param}.  Finally,
we use the NNLO MSTW2008 PDF set ({\tt MSTW8nn}) that corresponds to $\alpha_s(M_Z)=0.11707$.
\begin{table}[t]
\begin{center}
\begin{tabular}{|l|l||l|l|}
\hline
 $m_Z$            &  91.1876 GeV             &    $\alpha(m_Z)$    & 0.0075563839\\
 $m_W$            & 80.398 GeV               &    $\sin^2\theta_w$ & 0.2226459 \\
 $m_H$            & 125 GeV                  &    $m_t$            & 172 GeV\\
 $\Gamma_Z$       &  2.4952 GeV              &    $g_w^2$          & 0.4264904 \\
 $\Gamma_W$       &  2.1054 GeV              &    $e^2$            &0.0949563 \\
 $G_F$            & 0.116639$\times 10^{-4}$ & & \\
\hline
\end{tabular}
\caption{Masses, widths, couplings and scales used in the calculation of all processes.  Note
that the value of $\alpha(m_Z)$ corresponds to $1/\alpha(m_Z)=132.3384323$.}
\label{param}
\end{center}
\end{table}

An overview of the processes that will be studied in detail in this paper is
shown in Table~\ref{processes}\footnote{In addition, we include a more
  limited study of the di-photon process}.  As well as detailing the default
choice of renormalization and factorization scales ($\mu_R$ and
$\mu_F$), this table also shows the corresponding cross-section up to
NNLO.  The NNLO cross-sections are written in the form,
\begin{equation}
\label{eq:sigmadef}
\sigma_{NNLO} = \sigma_{LO} \times \left( 1 + \frac{\Delta\sigma_{NLO}}{\sigma_{LO}}
  + \frac{\Delta\sigma_{NNLO}}{\sigma_{LO}} \right) \,,
\end{equation}
so that, for instance, the corresponding NLO result is obtained by simply omitting the 
final term in this equation.  The cross-sections have been obtained by running the
readily-available public codes referenced in the final column of Table~\ref{processes}.
\begin{table}[t]
\begin{center}
\begin{tabular}{|l|l|l|l|l|}
\hline
Process     & $\mu_R$      & $\mu_F$      & Cross-section to NNLO & Reference \\
\hline
$gg \to H$  & $M_H$        & $M_H$        & $12.937\times(1+1.28+0.77) $ pb &{\tt ggh@nnlo}~\cite{Harlander:2003ai} \\
$Z$         & $2M_Z$       & $M_Z/2$      & $44.303\times(1+0.22+0.05) $ nb &{\tt ZWMS}~\cite{Hamberg:1990np} \\
$W^+$       & $2M_W$       & $M_W/2$      & $81.561\times(1+0.23+0.06) $ nb &{\tt ZWMS}~\cite{Hamberg:1990np} \\
$ZH$        & $\sqrt{q^2}$ & $\sqrt{q^2}$ & $0.68255\times(1+0.16+0.10)$ pb &{\tt vh@nnlo}~\cite{Brein:2003wg,Brein:2012ne} \\
$W^+H+W^-H$ & $\sqrt{q^2}$ & $\sqrt{q^2}$ & $1.2593\times(1+0.16+0.02) $ pb &{\tt vh@nnlo}~\cite{Brein:2003wg,Brein:2012ne} \\
\hline
\end{tabular}
\caption{Inclusive results from validation codes (listed in the final column) for processes considered in this paper.
NLO and NNLO corrections are shown as relative enhancements to the LO cross-section. $q^2$ is the overall invariant mass
squared of the vector boson and the Higgs boson.}
\label{processes}
\end{center}
\end{table}

We now describe the calculational setup that we use for these processes, which corresponds to the
default behavior of the above codes.  This behaviour has been matched in the MCFM code and, in
order to establish the equivalence of the parameters for MCFM and the other publicly available
codes, we compare results up to NLO in Table~\ref{LOandNLOcomparison}.
The agreement is excellent for all processes, so that we can be sure
that MCFM should produce the same results as the other codes when computing
the NLO and NNLO predictions using the $N$-jettiness subtraction method.

\subsection{Higgs production through gluon fusion}

We work in a theory in which only the top quark has a non-zero Yukawa
coupling.  Taking the large $m_t$ limit we obtain an effective Lagrangian that
expresses the coupling of gluons to 
the Higgs field~\cite{Wilczek:1977zn},
\begin{equation} \label{EffLag}
\mathcal{L}_H^{\mathrm{int}} = \frac{C(m_t^2,\mu^2)}{2} \, H\,\sum_a
G^a_{\mu\nu}\,G_a^{\mu\nu}\, . 
\end{equation}
where the sum is over the color degrees of freedom of the gluon. 
At the order required in this paper, the coefficient $C(m_t^2,\mu^2)$  
is given in the $\overline{\rm MS}$ scheme
by~\cite{Kramer:1996iq,Chetyrkin:1997iv},
\begin{eqnarray}\label{Ct}
   C(m_t^2,\mu^2) &=& \frac{\alpha_S}{6 \pi v} \Bigg\{
    1 + \frac{\alpha_s}{4\pi}\,(5C_A-3C_F) \\
   &&\mbox{}+ \left( \frac{\alpha_s}{4\pi} \right)^2
    \bigg[ \frac{27}{2}\,C_F^2 
    + \left( 11\ln\frac{m_t^2}{\mu^2} - \frac{100}{3} \right) C_F C_A 
    - \left( 7\ln\frac{m_t^2}{\mu^2} - \frac{1063}{36} \right) C_A^2 
    \nonumber\\
   &&\quad\mbox{}- \frac{4}{3}\,C_F T_F - \frac{5}{6}\,C_A T_F 
   - \left( 8\ln\frac{m_t^2}{\mu^2} + 5 \right) C_F T_F n_f 
   - \frac{47}{9}\,C_A T_F n_f \bigg] \Bigg\} \,. \nonumber
\end{eqnarray} 
Here $v$ is the vacuum expectation value of the Higgs field, $v = 246$ GeV. 
The only remaining $m_t$-dependence at this order is the one shown in the
$O(\alpha_s^2)$ contribution to the coefficient of the effective operator.

The validation cross-section for this process is obtained using {\tt ggh@nnlo}~\cite{Harlander:2003ai}.
As can be seen from Table~\ref{processes}, at $13$~TeV the higher-order corrections to
this cross-section are quite large.


\subsection{$W$ and $Z$ production}

\begin{table}[t]
\begin{center}
\begin{tabular}{|l|l|l|l|c|}
\hline
Process        & Order    & MCFM  cross-section    & Cross-check \\
\hline  
$H$ production & LO       & $12.937 \pm 0.001$  pb & 12.937 pb \\
               & NLO      & $29.520 \pm 0.001$  pb & 29.521 pb \\
\hline
$Z$            & LO       & $44.303 \pm 0.001$  nb & 44.303 nb \\
               & NLO      & $53.958 \pm 0.002$  nb & 53.957 nb \\
\hline
$W^+$          & LO       & $81.559 \pm 0.002$  nb & 81.561 nb \\
               & NLO      & $100.298 \pm 0.003$ nb & 100.299 nb \\
\hline
$ZH$           & LO       & $0.68254 \pm 0.00001$   pb & 0.68255 pb \\
               & NLO      & $0.79073 \pm 0.00003$   pb & 0.79079 pb \\
\hline
$W^+H+W^-H$    & LO       & $1.2592 \pm 0.02$   pb & 1.2593 pb \\
               & NLO      & $1.4629 \pm 0.04$   pb & 1.4630 pb \\
\hline
\end{tabular}
\caption{\label{LOandNLOcomparison}
Comparison of LO and NLO cross-sections computed using the standard MCFM
subtraction method with the codes used for cross-checking in this paper.}
\end{center}
\end{table}
 
To establish the correct values of the higher-order cross sections for
$W^+$ and $Z$ production we use the program {\tt ZWMS}~\cite{Hamberg:1990np}.
For the sake of illustration we have chosen to perform the comparison for only one
charge of the $W$-boson.
We note that for the canonical scale choice $\mu_R = \mu_F = M_V$ (where $V=W, Z$)
the NNLO corrections  are very small.
Although this is ultimately an advantage in
terms of the accuracy required for phenomenological applications, it prohibits
a careful study of the behaviour of the $N$-jettiness calculation.  To enhance
the size of the NNLO correction we therefore use an asymmetric 
choice,  $\mu_R = 2M_V$, $\mu_F=M_V/2$.  This results in NNLO corrections of approximately $5$\% relative to
the LO cross-section at 13 TeV (c.f. Table~\ref{processes}).

Note that by default MCFM includes the decay of the vector bosons, $Z/\gamma^*$ to a lepton pair.
For comparison with the rate for production of on-shell $Z$-bosons, we remove the (small) contribution
mediated by a virtual photon and divide out the overall branching ratio of the $Z$-boson to leptons.



\subsection{Associated Higgs production: $W^\pm H$ and $ZH$ processes}

To establish target NLO and NNLO values for the total cross section for
$W^\pm H$ and $ZH$ production we use the program {\tt vh@nnlo}~\cite{Brein:2003wg,Brein:2012ne}.
In order to facilitate an easy comparison with this program, we use the scale choice
$\mu_R = \mu_F = \sqrt{q^2} \equiv \sqrt{(p_V + p_H)^2}$, with $V=W^\pm$ or $V=Z$ as appropriate.
We also sum over both charges of the $W$ boson, i.e. we include both $W^+$ and $W^-$ contributions
-- which can differ substantially at a $pp$ collider such as the LHC -- in all of the results below.
For the diagrams in which the Higgs boson couples directly to a top quark loop
we work in the effective theory, valid in the large $m_t$ limit given
by Eq.~(\ref{EffLag}). A detailed phenomenological study of the NNLO implementation of these processes
in MCFM has been presented in Ref.~\cite{Campbell:2016jau}.

For both $W^\pm H$ and $ZH$ processes the correction originating
from diagrams with the Higgs boson coupling to a top quark loop
is approximately $1.5\%$.   The $ZH$ process also includes 
a substantial finite component due to $gg \to ZH$ loops at NNLO.  
The NNLO corrections that correspond to simple dressings of the
LO diagrams are very small, of order $1\%$, for both $W^\pm H$ and $ZH$ 
production.  The net effect of all these contributions is shown
in Table~\ref{processes}, where the NNLO corrections to the $ZH$ process
are at the level of $10$\%.  In contrast,  the total
NNLO correction to the $W^\pm H$ process is about $2$\% of
the LO result.

\subsection{Diphoton production}
NNLO predictions for the diphoton process, obtained using MCFM, have been
presented in ref.~\cite{Campbell:2016yrh}.  Therein the results have been
validated using the same procedure as we will adopt later;  we do not repeat
that analysis here.  However, we will later on summarize the size of the
power corrections and timing results for this process.

\section{$N$-Jettiness subtraction at NLO}
\label{sec:nlo}
\begin{table}[t]
\begin{center}
\begin{tabular}{|l|r|c|}
\hline
Process     & $\Delta\sigma^{NLO}$ & $\Delta\sigma^{NLO}/\sigma^{LO}$ \\
\hline
$H$         & 16.58  pb            & 1.282 \\
$Z$         & 9.655  nb            & 0.218 \\
$W^+$       & 18.74  nb            & 0.230 \\
$ZH$        & 0.1082 pb            & 0.158 \\
$W^+H+W^-H$ & 0.2037 pb            & 0.162 \\
\hline
\end{tabular}
\caption{\label{NLOcorrections}
NLO corrections to the processes computed in this paper using the
$N$-jettiness method.}
\end{center}
\end{table}

Although the calculation of NLO corrections for the processes considered here is straightforward,
a detailed examination of the corresponding $N$-jettiness subtraction calculation is extremely
useful.  It provides a stringent check of the accuracy of this approach, namely a direct probe of the
size of the power corrections that have been neglected in Eq.~(\ref{eq:SCETfac1}).  This can be tested with
exquisite accuracy, due to the relative simplicity of the calculation compared to the corresponding
exercise at NNLO.  This comparison can also illuminate the limitations of this approach when moving
beyond an inclusive calculation, by using MCFM to compare calculations of differential distributions
at NLO.

The calculation of NLO corrections using the $N$-jettiness subtraction method
is straightforward in MCFM.  The below-cut contribution is easily computed,
while the above-cut contribution corresponds to a LO calculation of the
process that contains an additional parton.  
In order to avoid numerical instability in calculations using MCFM,
previous versions of the code have applied a small cutoff on all invariant
masses present in the problem, $\sqrt{s_{ij}} > {\tt cutoff}$.  In this version
this has been changed so as to enforce a small cutoff on the partonic jettiness
of every parton present in a given calculation, $\taun(p_j) > {\tt cutoff}$.
Since the above-cut region involves a standard LO calculation, for which there
are no numerical instabilities, we are able to choose a value for this cutoff
close to the limit of double precision, ${\tt cutoff}=10^{-12}$ GeV.

\subsection{Inclusive cross-sections}

The benchmark cross-sections that form the basis for this comparison
can be extracted from Table~\ref{LOandNLOcomparison} and, for convenience,
have been summarized in Table~\ref{NLOcorrections}.  As is well-known,
the NLO corrections to Higgs production through gluon fusion are very large,
while all of the other processes receive corrections of order $20$\%.

\begin{figure}[t]
\centering
\includegraphics[width=11.cm]{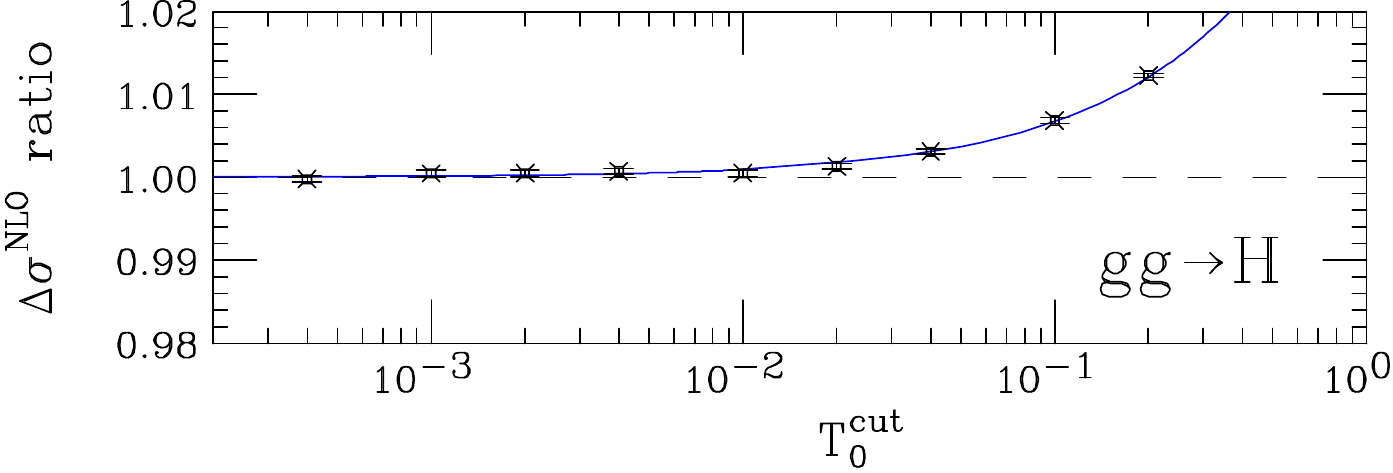} \\
\includegraphics[width=11.cm]{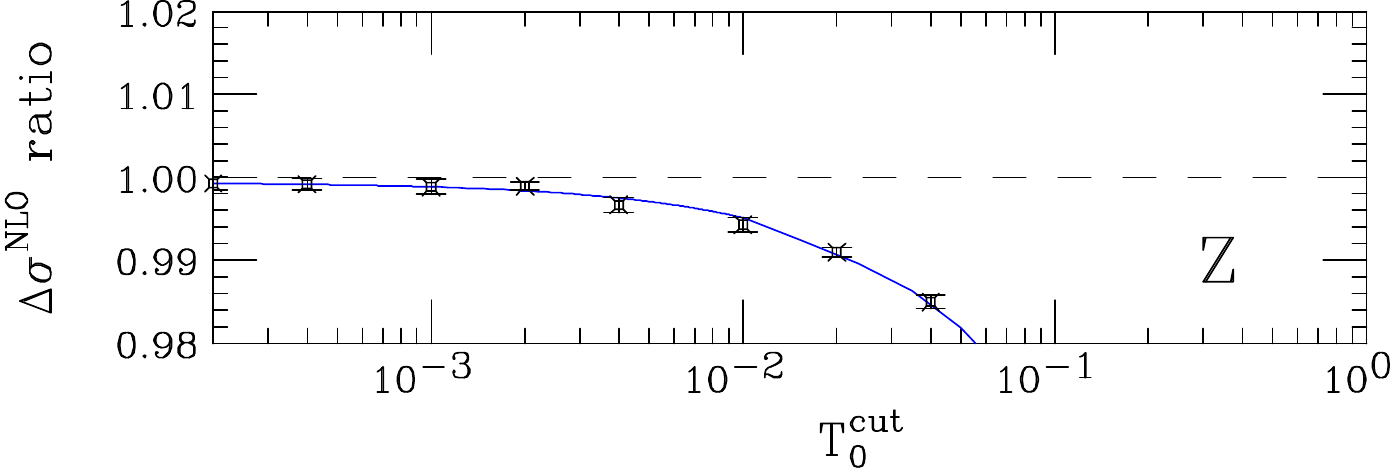} \\
\includegraphics[width=11.cm]{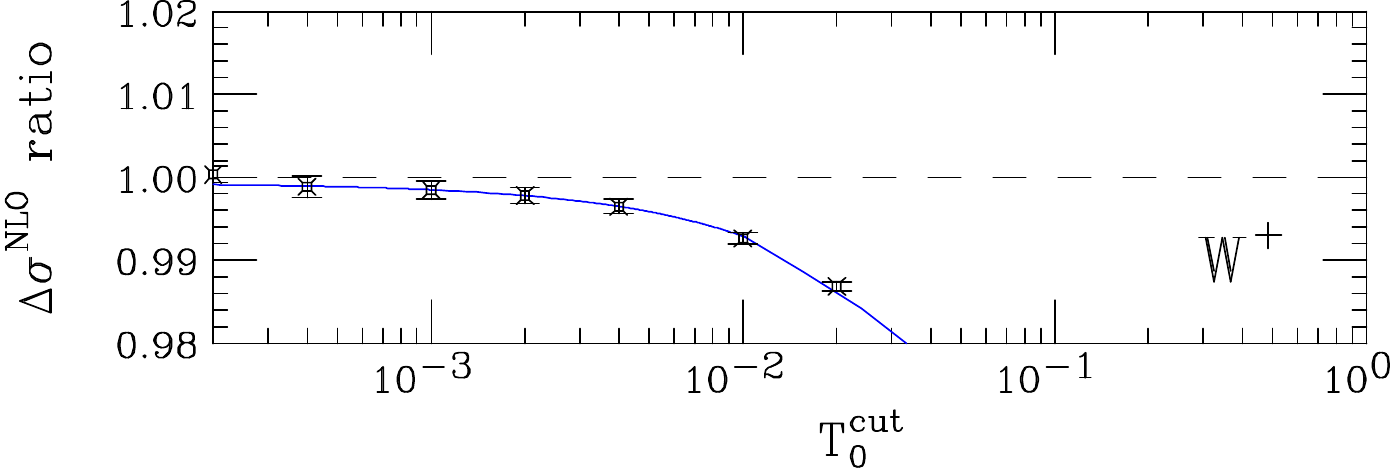} \\
\includegraphics[width=11.cm]{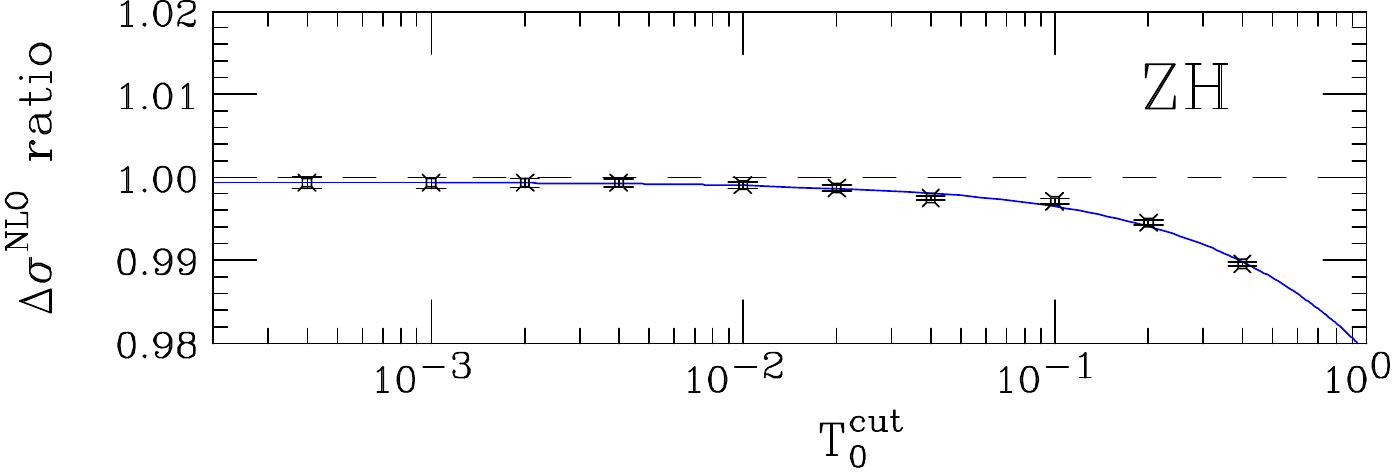} \\
\includegraphics[width=11.cm]{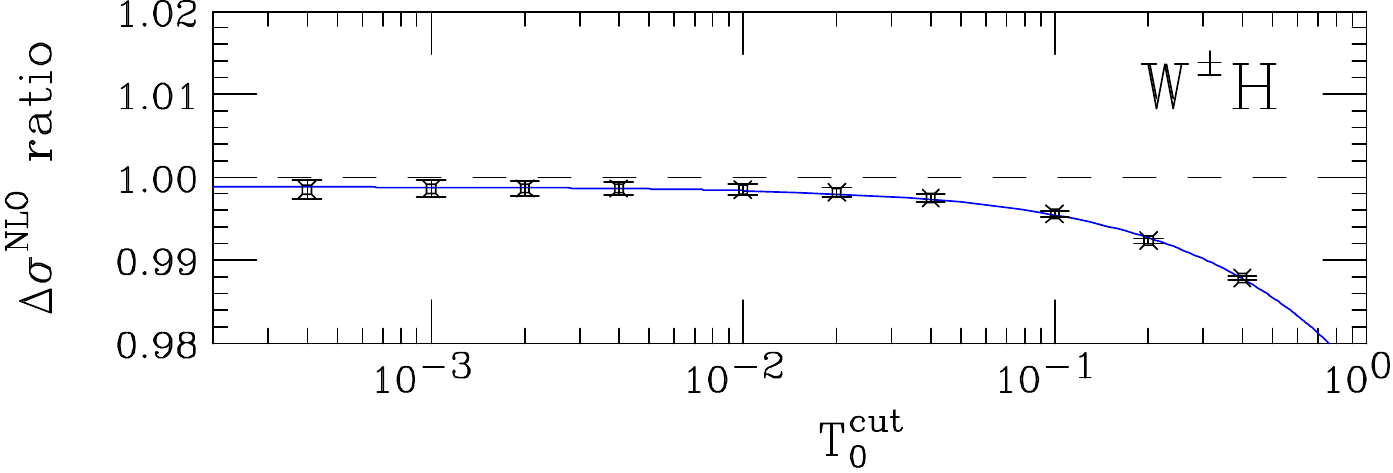}
\caption{\label{taudepnlo} The ratio of the NLO correction calculated
using $N$-jettiness subtraction as implemented in MCFM to the standard MCFM
subtraction result (as presented in Table~\ref{NLOcorrections}).  The
ratios are plotted as a function of the $N$-jettiness resolution parameter
$\tauzerocut$ in GeV.  The comparison is performed for $gg \to H$,
$Z$, $W^+$, $ZH$ and $W^\pm H$ production and the lines represent
fits to the individual points using the form given in
Eq.~(\ref{eq:nloform}).}
\end{figure}

A comparison of the $N$-jettiness calculations of these coefficients, with the results
shown in Table~\ref{NLOcorrections}, is shown in Fig.~\ref{taudepnlo}.  The ratio
of the calculations is shown as a function of $\tauzerocut$, for a range
of suitable values of $\tauzerocut$.  The approach of the $N$-jettiness
calculation to the correct result as $\tauzerocut \to 0$ is clear for
each process.  However the manner in which the correct result is reached varies considerably.  For instance,
Higgs production through gluon fusion approaches the correct result from above, while
the other processes approach it from below.   The approach is much slower for
$W^+$ and $Z$ production than for any of the Higgs production processes, with percent
level accuracy only reached for $\tauzerocut \lesssim 0.01$ GeV.

The approach of the $N$-jettiness subtraction result to the correct answer
is determined by the behaviour of power corrections that are not
accounted for at present.  At the NLO level after integration over the final-state phase space this can be modeled by the following functional form,
\begin{equation}
\Delta \sigma^{NLO}_{\mbox{\tiny jettiness}} (\tauzerocut) =
  \Delta\sigma^{NLO} + c\times\left(\frac{\tauzerocut}{Q}\right)\times\log\left(\frac{\tauzerocut}{Q}\right) \;,
\label{eq:nloform}
\end{equation}
where $Q$ is the appropriate scale for the process at hand and $c$ is
an unknown constant.  For single boson production $Q$ is taken to be
the mass of the produced particle ($M_H$, $M_W$ or $M_Z$) while for
the associated production processes we use $Q=M_W + M_H$ and $Q=M_Z +M_H$.  
Fig.~\ref{taudepnlo} also shows a fit of the results to
Eq.~(\ref{eq:nloform}), with the values of $\Delta\sigma^{NLO}$ and
$c$ determined in the fit. The difference of the fit value for
$\Delta\sigma^{NLO}$ with the known result given in
Table~\ref{NLOcorrections} is no larger than one per mille for all
processes.  The results of a study of the dominant NLO power correction, obtained using an analytic calculation~\cite{BPL},
will be discussed in Section~\ref{power}.
 
Since the speed of the approach to the correct result is qualitatively
much worse for $W$ and $Z$ production it is instructive to examine the
processes in more detail in order to uncover the origin of the
difference.  To that end we now turn to a comparison of more
differential results.

\subsection{Rapidity distributions at NLO}

The simplest distribution to study is the rapidity of the produced system, which is intimately related to the
momentum fractions carried by the incident partons.  We will compare the prediction for the NLO contribution to
this distribution (i.e. corresponding to $\Delta\sigma^{NLO}$) computed using dipole subtraction and jettiness
subtraction with $\tauzerocut=0.01$ GeV and $\tauzerocut=0.04$ GeV.  The difference between
the true result and the jettiness calculation for $\tauzerocut=0.04$
GeV is about $0.4$\% for $gg \to H$,
$1.5$\% for $Z$ production and $0.3$\% for $ZH$.  These processes are sufficient to illustrate the issue, since 
$W^+$ and $W^\pm H$ production show very similar behaviour to the $Z$ and $ZH$ processes respectively.

\begin{figure}[t]
\centering
\includegraphics[width=0.32\textwidth]{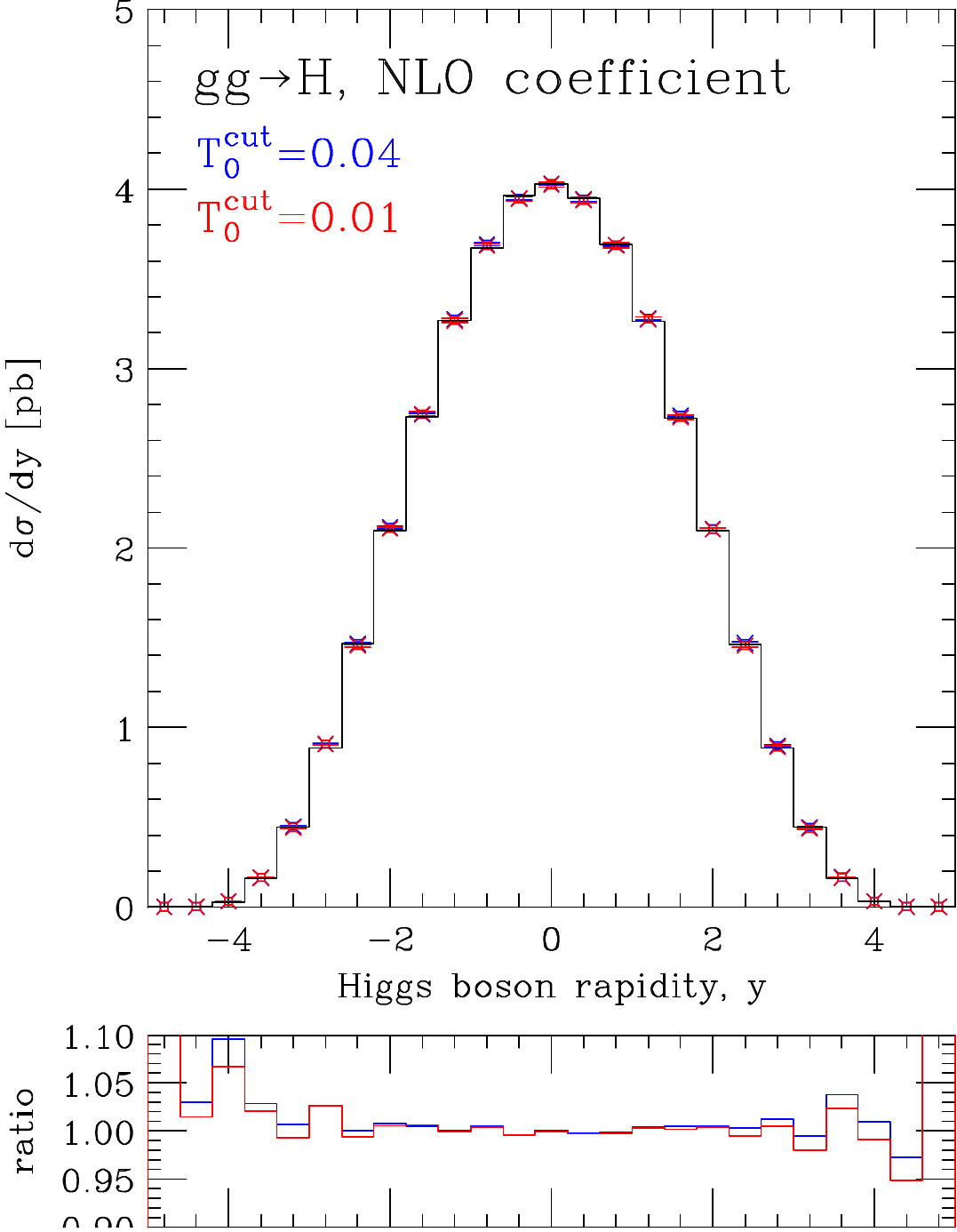}
\includegraphics[width=0.32\textwidth]{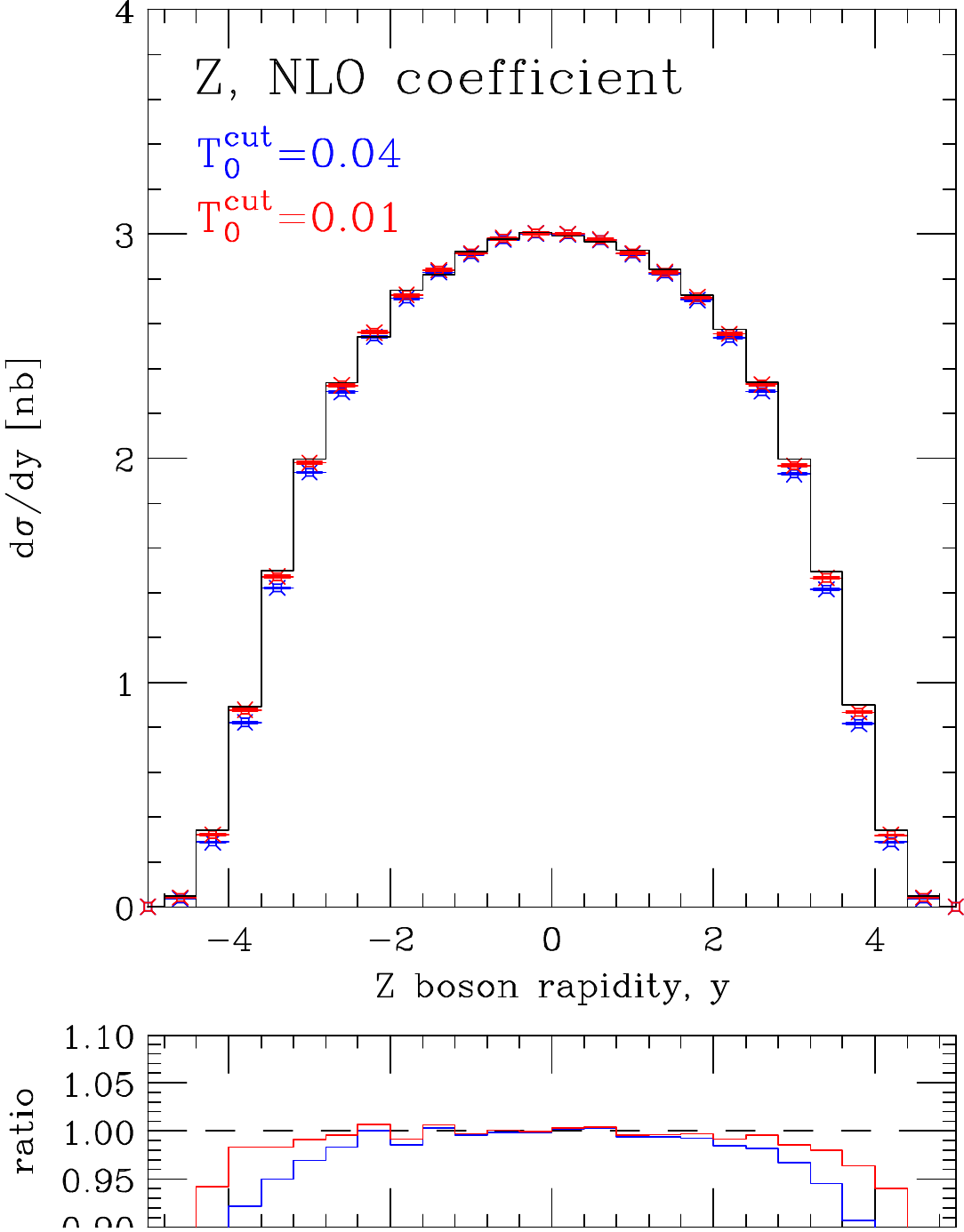}
\includegraphics[width=0.32\textwidth]{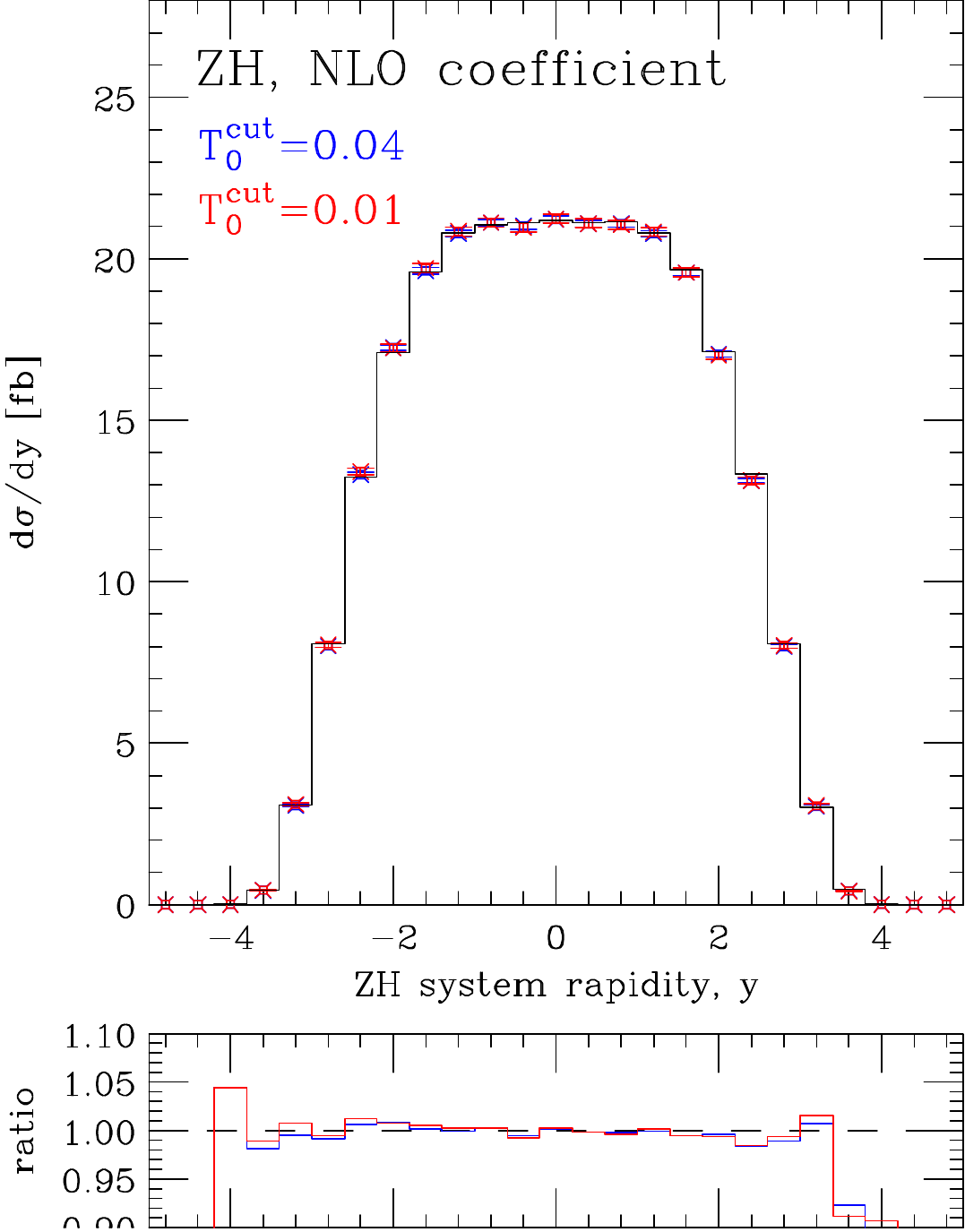}
\caption{\label{ynlo}MCFM calculation using $N$-jettiness subtraction of the NLO contribution to the rapidity distributions
of the Higgs boson (left), $Z$ boson (centre) and $ZH$ system (right), in the $gg \to H$, $Z$ and $ZH$ production
processes respectively. Results are shown 
for two values of $\tauzerocut$ (in GeV) and are compared with the normal MCFM calculation (solid histogram).
The lower panel shows the ratio of the jettiness results to the normal MCFM calculation. }
\end{figure}

Results are shown in Fig.~\ref{ynlo}.  The agreement of the jettiness calculations with the normal MCFM result is
excellent overall, particularly for central production $|y| \lesssim 3$.  However there is evidence for an increase
in the size of the power corrections at larger absolute
rapidities\footnote{This can be expected as $\tau_0\sim p_T\exp(-|\eta|)$ and
therefore at large rapidity the transverse momentum cutoff is larger,
increasing the expected power corrections}.   The reason for the qualitative difference in the
behaviour is thus two-fold.  First, the onset of power corrections with increased rapidity occurs sooner for $Z$
production.  Second, and critically, the shape of the rapidity distribution is much broader for $Z$ production so that
the effect of the high-rapidity tails is more apparent in the inclusive rates presented in the previous
section.   It suggests that a restriction to more central rapidities would decrease the effect of power corrections
and speed the convergence to the correct result. 

\subsection{Cross-sections under cuts}

As an explicit demonstration of this behavior we will contrast the
effect of the power corrections on the inclusive cross-section with
the behavior under a more realistic set of experimental cuts.  Rather
than cutting directly on the rapidity of the $W$ or $Z$ boson, we
instead apply a minimal set of cuts on the $W$ and $Z$ boson decay
products that might be applied in an experimental analysis.  We
consider a $Z$ boson decay to an electron-positron pair and demand
that both leptons be observed in the central region, $|y(e^\pm)|<2.5$.
For the $W^+$ boson case we consider the decay into an positron and
neutrino, imposing a rapidity constraint on the charged lepton
$|y(e^+)|<2.5$ and a minimum missing transverse energy (MET) of 30
GeV.  Note that the application of these cuts means that a comparison
with the code {\tt ZWMS} can no longer be made.  Although
DYNNLO~\cite{Catani:2009sm} or FEWZ~\cite{Gavin:2010az,Li:2012wna} could be used to
provide a reference cross-section under these cuts we do not pursue
that here.  Instead we simply normalize to the (fitted) asymptotic
result.

\begin{figure}[t]
\centering
\includegraphics[width=11.cm]{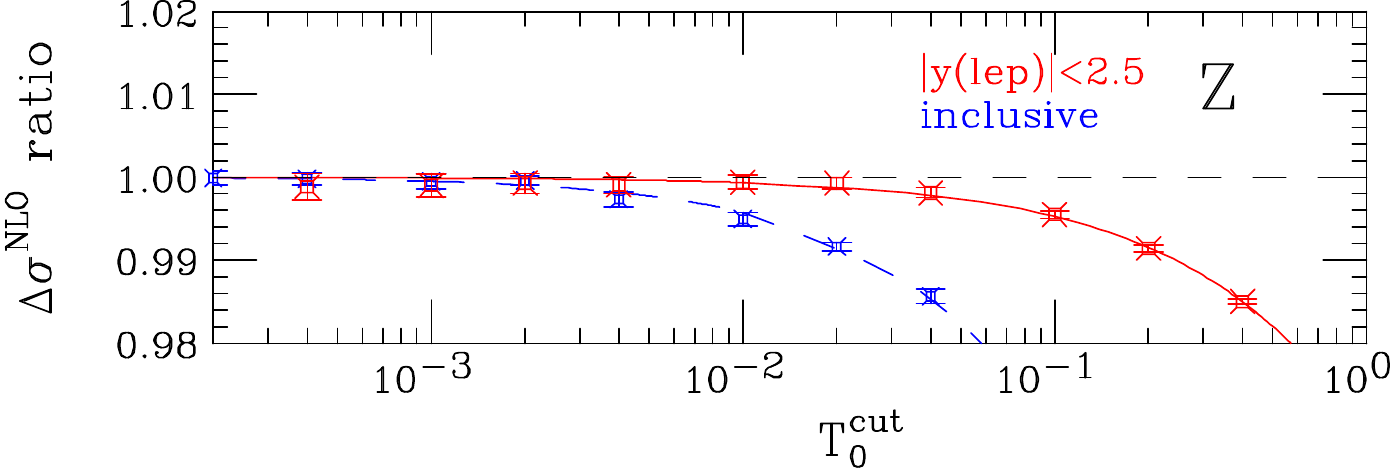} \\
\includegraphics[width=11.cm]{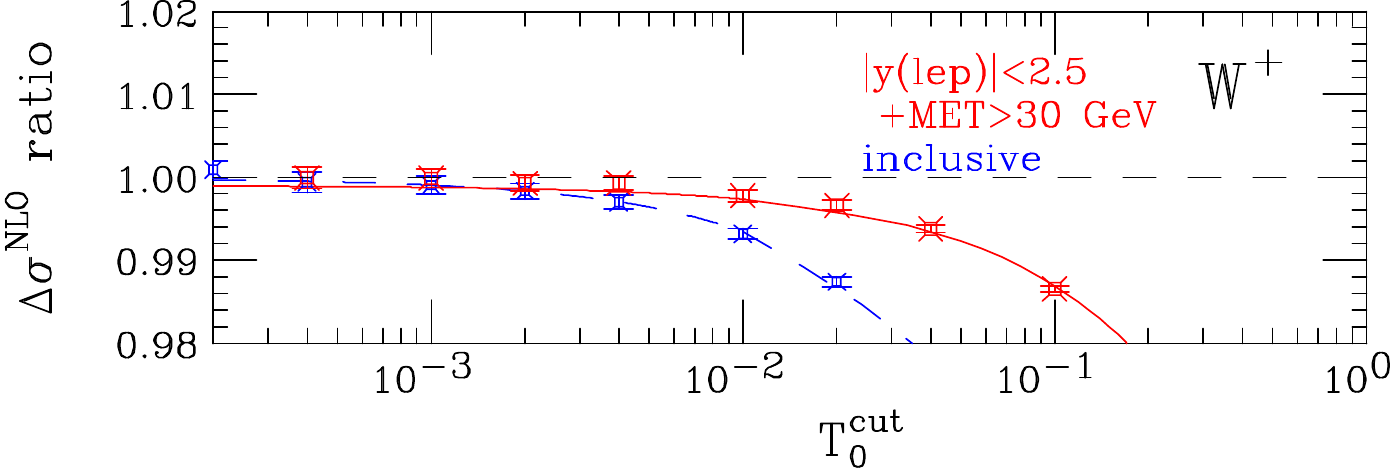} \\
\caption{\label{taudepnlofid}The ratio of the MCFM $N$-jettiness calculation of the NLO coefficient
to the (fitted) asymptotic result, as a function of the $N$-jettiness
resolution parameter $\tauzerocut$ in GeV.  The comparison is performed for
$Z$ (top) and $W^+$ production (bottom) and for both the inclusive case and for a minimal
set of fiducial cuts (detailed in the text). The lines represent fits to the individual
points using the form given in Eq.~(\ref{eq:nloform}).}
\end{figure}

The results of this study are shown in Fig.~\ref{taudepnlofid}. As anticipated, the effect
of the cuts is to significantly decrease the $\tauzerocut$-dependence of
the cross-section.  For instance, rather than a difference of approximately $1$\% with
the asymptotic result for $\tauzerocut = 0.02$ GeV in the inclusive case, the
fiducial cross-section differs by a few per mille or less for the same value of
$\tauzerocut$.  The inability to restrict the rapidity of the unobserved
neutrino in the case of $W^+ \to e^+ \nu$ production, compared to $Z \to e^- e^+$, leads
to a slightly slower approach to the correct result. 

\section{$N$-Jettiness subtraction at NNLO}
\label{sec:nnlo}
At NNLO, the $N$-jettiness subtraction method involves an above-cut contribution
that corresponds to a NLO calculation of the process containing an additional parton.
In contrast to the previous order, this results in genuine numerical instabilities that
primarily arise from the cancellation of
subtraction terms in the real radiation contribution.
As a result we must use a larger value of the safety-cutoff parameter, namely
${\tt cutoff}=10^{-8}$ GeV.  This is appropriate for computations
in double precision, such as the ones presented in this paper. Although we do not
include any quadruple precision results here, we note that this cut may be relaxed
significantly in that case.  We note the caveat that the running time of the code
increases significantly in quadruple precision, by about an order of magnitude.

\subsection{Inclusive cross-sections}

The expected NNLO cross-sections in the inclusive case, obtained using the
already-available public codes listed previously, are shown in Table~\ref{NNLOcorrections}.
The corrections to the $gg \to H$ process are again large at this order, while all
of the other processes have corrections in the $2$--$10$\% range.
Of these other processes
$ZH$ production has the largest correction, but this is largely due to the effect of
finite $gg \to ZH$ and top-Yukawa contributions, as discussed previously.
\begin{table}[t]
\begin{center}
\begin{tabular}{|l|r|c|}
\hline
Process     & $\Delta\sigma^{NNLO}$ & $\Delta\sigma^{NNLO}/\sigma^{LO}$ \\
\hline
$H$         & 10.01   pb 	   & 0.774 \\
$Z$         & 2.200   nb 	   & 0.050 \\
$W^+$       & 4.702   nb 	   & 0.058 \\
$ZH$        & 0.06562   pb 	   & 0.096 \\
$W^+H+W^-H$ & 0.0294    pb          & 0.023 \\
\hline
\end{tabular}
\caption{\label{NNLOcorrections}
NNLO corrections to the processes computed in this paper.}
\end{center}
\end{table}

\begin{figure}[t]
\centering
\includegraphics[width=11.cm]{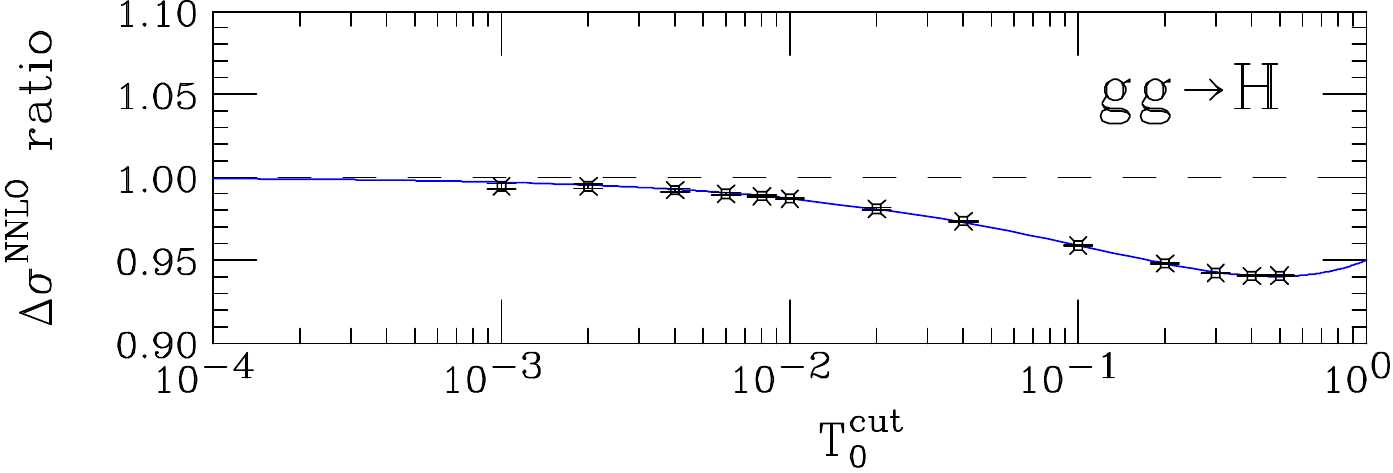} \\
\includegraphics[width=11.cm]{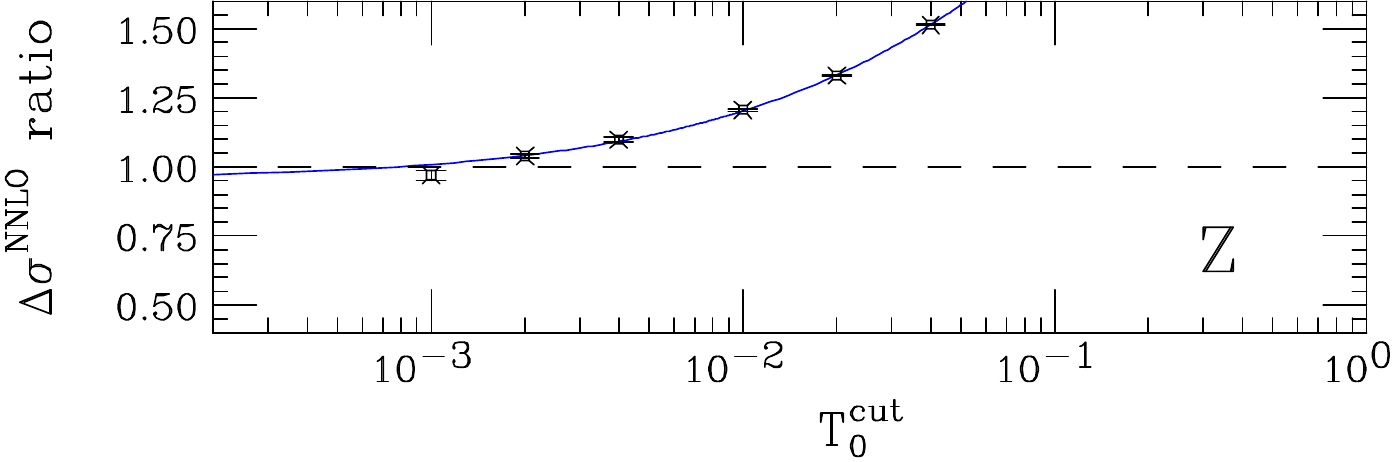} \\
\includegraphics[width=11.cm]{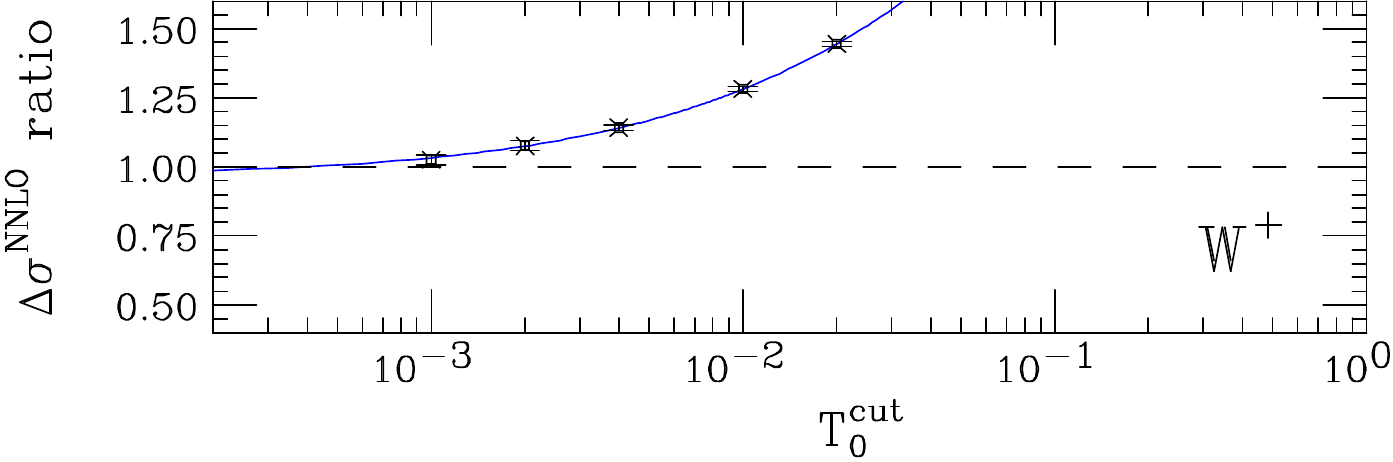} \\
\includegraphics[width=11.cm]{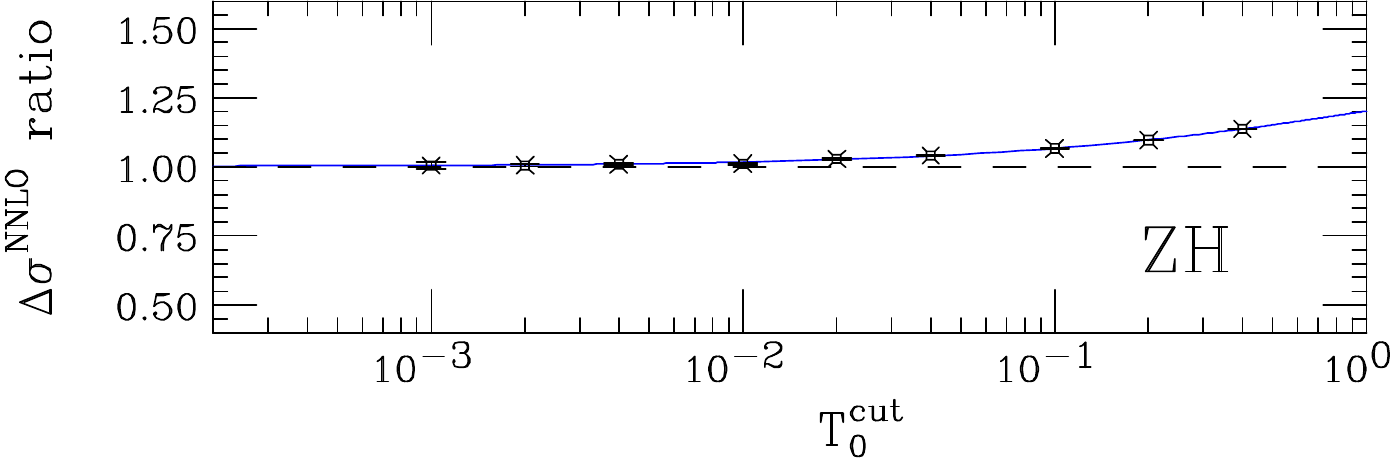} \\
\includegraphics[width=11.cm]{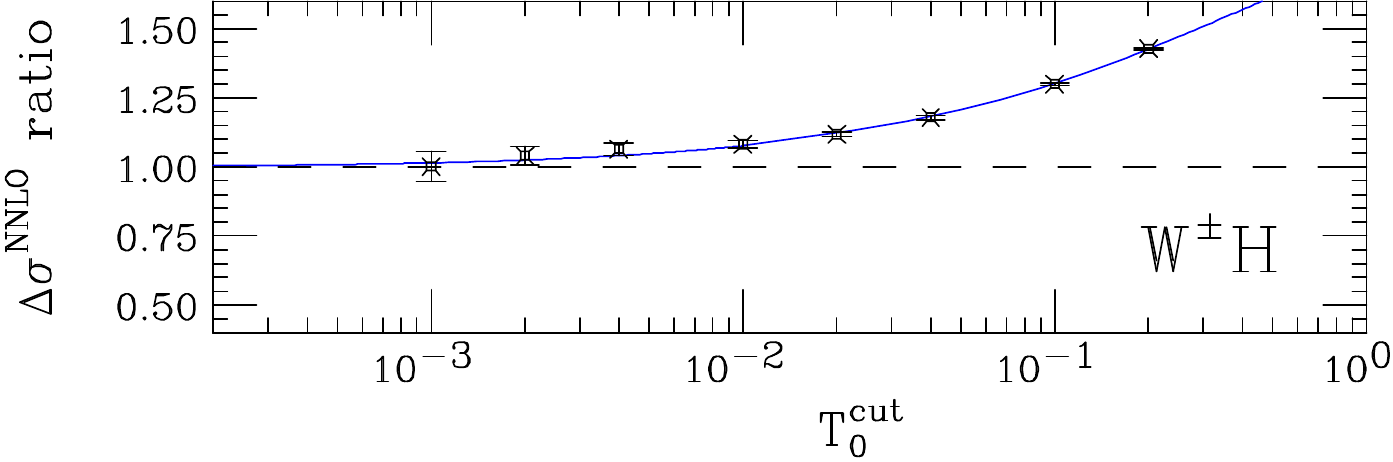}
\caption{\label{taudepnnlo}The ratio of the MCFM calculation of the NNLO coefficient
to the known result presented in Table~\ref{NNLOcorrections}, as a function of the $N$-jettiness
resolution parameter $\tauzerocut$ (in GeV).  The comparison is performed for
$gg \to H$, $Z$, $W^+$, $ZH$ and $W^\pm H$ production and
the lines represent fits to the individual points using the form given in Eq.~(\ref{eq:nnloform}).}
\end{figure}

The calculation of the NNLO coefficients by jettiness subtraction are compared with
results from the literature in Fig.~\ref{taudepnnlo}.  Note that all of the plots use a common scale
for the ordinates, which display the ratio, except for the one representing the $gg \to H$ calculation, for which the power corrections
are much smaller.
It is clear from this figure that there is a slower approach to the asymptotic result
than at NLO, but that excellent agreement is still obtained for smaller values of $\tauzerocut$.
The relatively poorer approach to the true result is expected from the behaviour of the power corrections
at NNLO, whose leading two terms can be modeled after integration over the final-state phase space as
\begin{equation}
\Delta \sigma^{NNLO}_{\mbox{\tiny jettiness}} (\tauzerocut) =
  \Delta\sigma^{NNLO}
  + c_3\times\left(\frac{\tauzerocut}{Q}\right)\times\log^3\left(\frac{\tauzerocut}{Q}\right)
  + c_2\times\left(\frac{\tauzerocut}{Q}\right)\times\log^2\left(\frac{\tauzerocut}{Q}\right)
   \;,
\label{eq:nnloform}
\end{equation}
where $Q$ is the appropriate scale as before and $c_{2,3}$ are unknown constants. 
Also shown in Fig.~\ref{taudepnnlo} are fits of the results to Eq.~(\ref{eq:nnloform}), with the
values of $\Delta\sigma^{NNLO}$ and $c_{2,3}$  determined in the fit.  The subleading term is only important
in the case of the $gg \to H$ process, in order to capture the observed turn-over for larger values of
$\tauzerocut$.
For $gg \to H$, $ZH$ and $W^\pm H$ production the fit value of $\Delta\sigma^{NNLO}$ differs from the
known result given in Table~\ref{NNLOcorrections} by less than one per mille.  For the $Z$ and $W^+$ processes
the agreement is not as good, at the level of approximately $4$\%.
Again, the dominant NNLO power correction can be calculated analytically from first principles~\cite{BPL}
and its impact will be shown in Section~\ref{power}.

\subsection{Rapidity distributions at NNLO}

Given the effect of the power corrections on the rapidity distribution at NLO, we expect to see a
similar pattern at NNLO.  We  compare predictions for $\tauzerocut=0.01$ GeV and
$\tauzerocut=0.004$ GeV.  For the $gg \to H$ and $ZH$ processes that we study here, the
predictions for $\tauzerocut=0.004$ GeV
should be a good proxy for the exact distribution given the small deviations from the inclusive cross-section
to which they correspond (around $0.8$\% for both).  For $Z$ production, this value of $\tauzerocut$
yields a total cross-section that differs by $10$\% from the known result.   To obtain an actual phenomenological result one must run with a lower $\tauzerocut$. Nevertheless it is sufficient to demonstrate
the pattern of the power corrections.

\begin{figure}[t]
\centering
\includegraphics[width=0.32\textwidth]{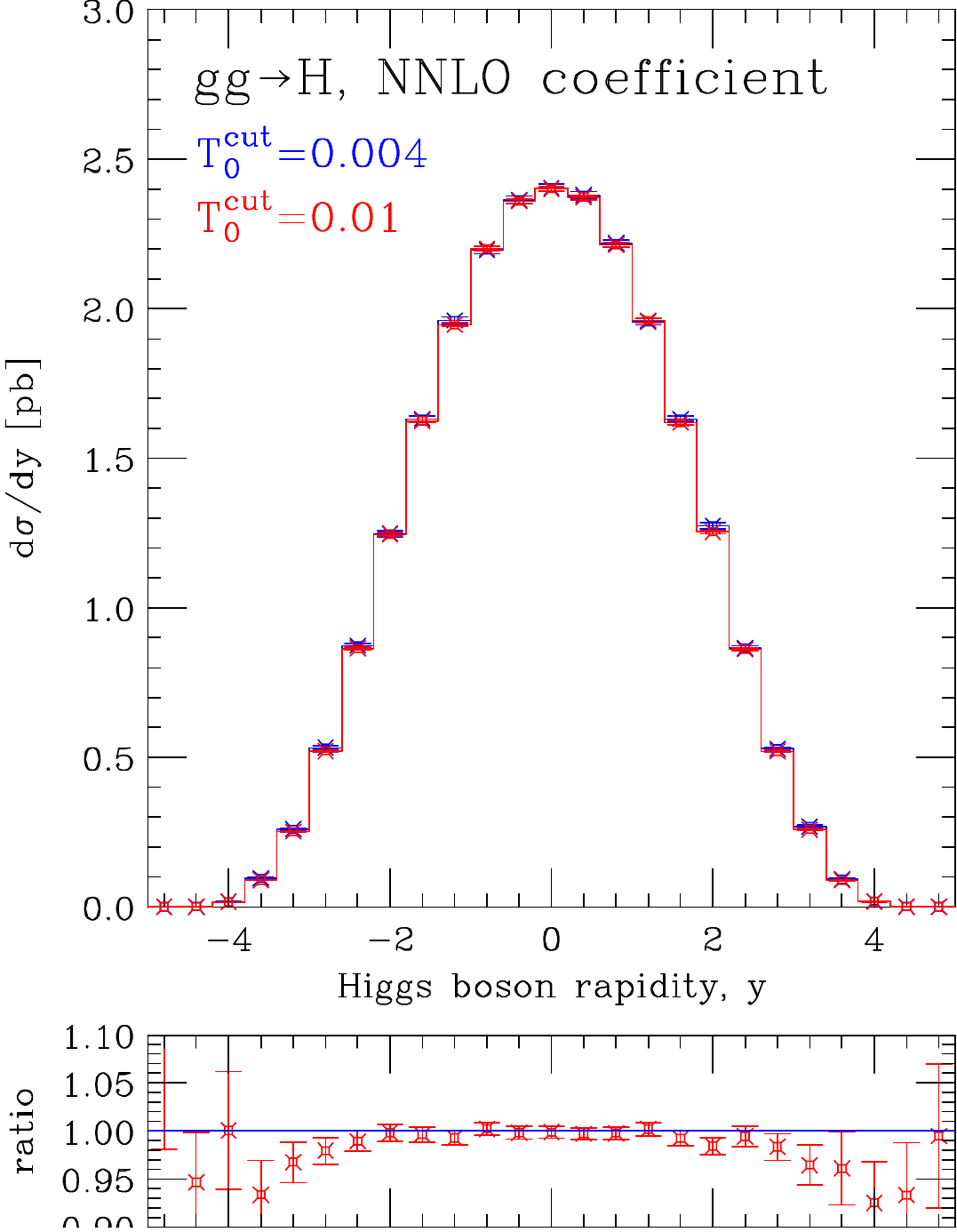}
\includegraphics[width=0.32\textwidth]{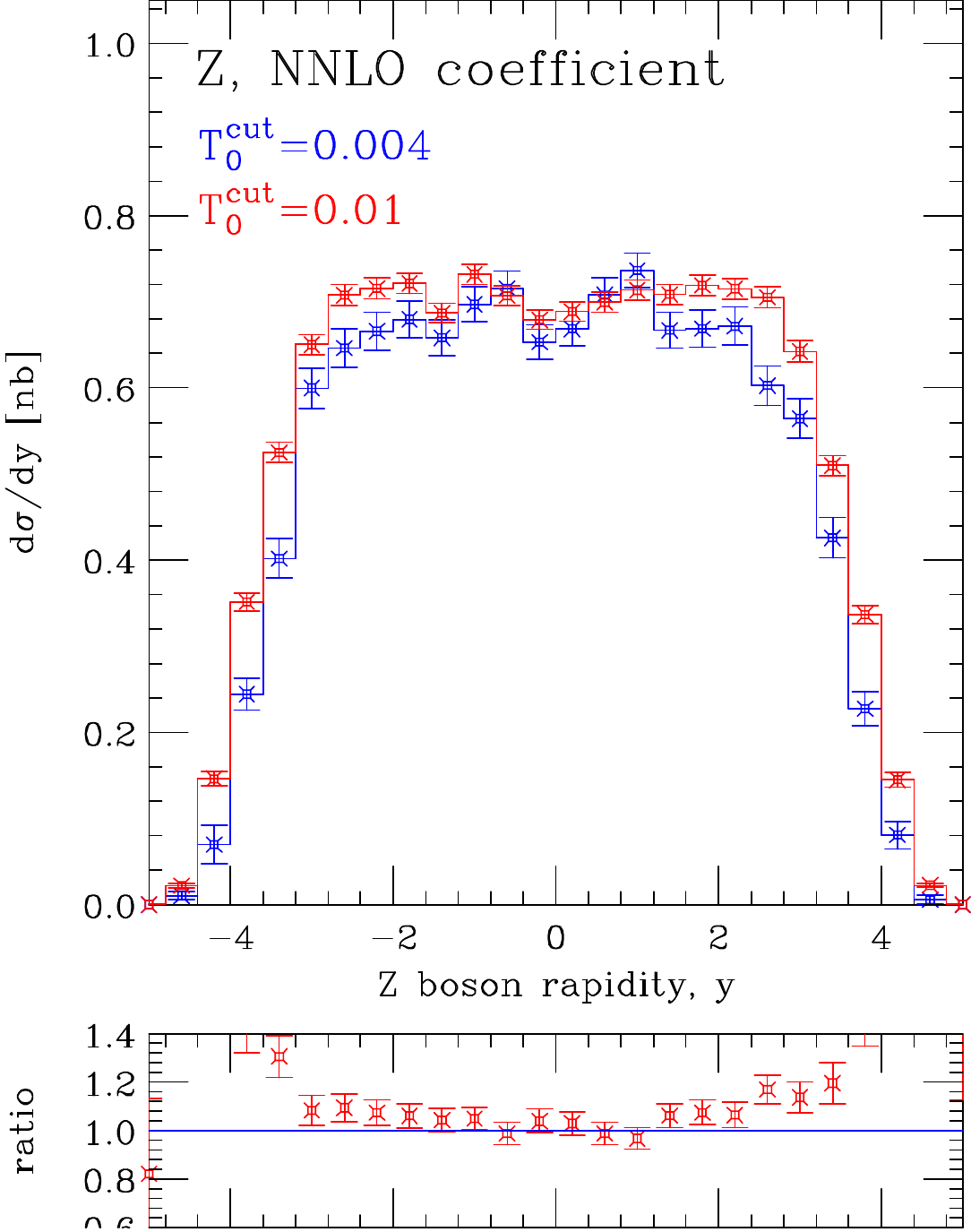}
\includegraphics[width=0.32\textwidth]{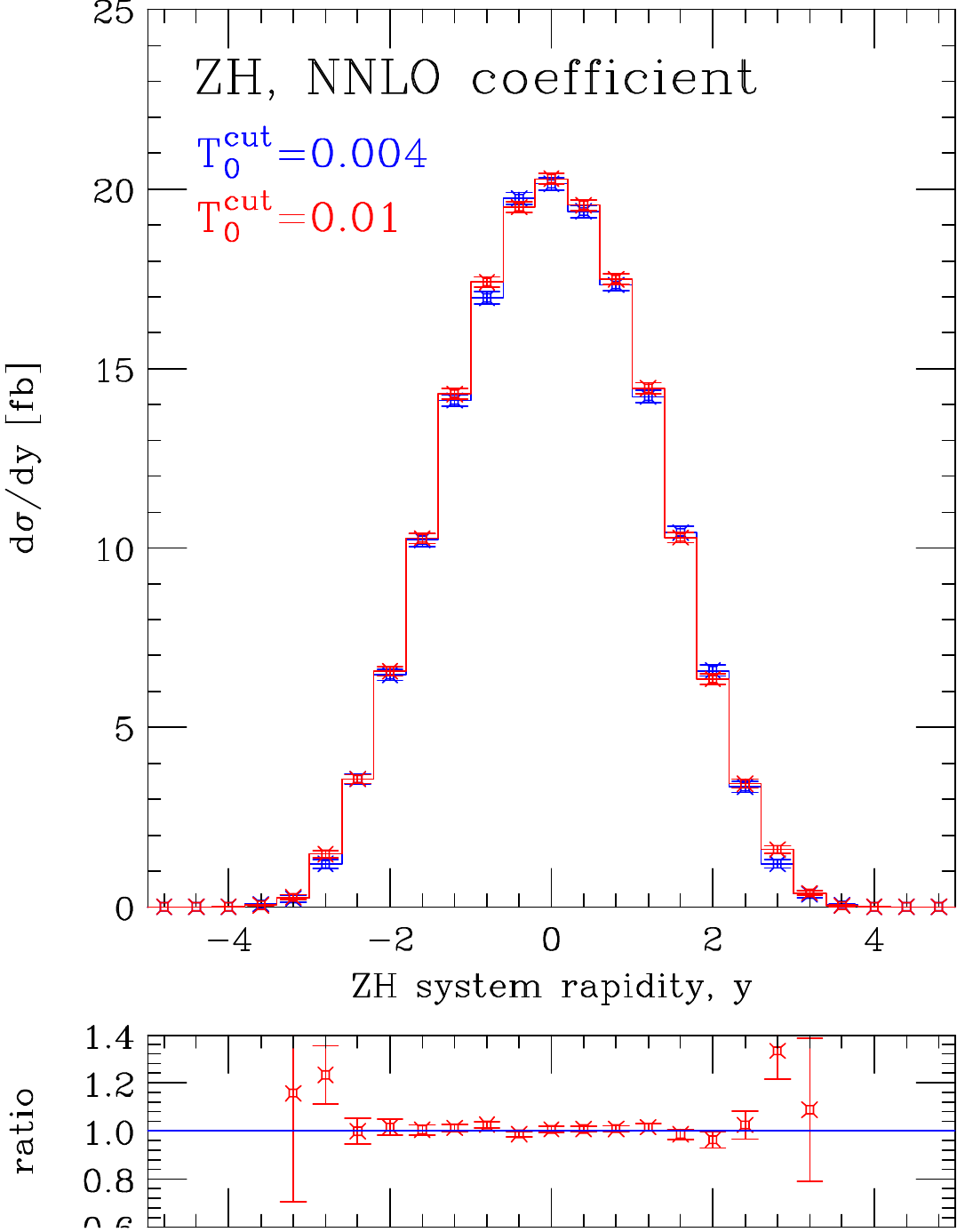}
\caption{\label{ynnlo}MCFM calculation of the NNLO contribution to the rapidity distributions
of the Higgs boson (left), $Z$ boson (centre) and $ZH$ system (right), in the $gg \to H$, $Z$ and $ZH$ production
processes respectively. Results are shown for two values of $\tauzerocut$, with the lower panel showing the ratio of
the $\tauzerocut=0.01$ GeV result to the $\tauzerocut=0.004$ GeV one.}
\end{figure}

The dependence on $\tauzerocut$ of the NNLO contributions to the rapidity distributions is
illustrated in Fig.~\ref{ynnlo}.  As observed at NLO, all three distributions are much less sensitive to the choice
of $\tauzerocut$ in the central region than at large rapidities.  The quality of the independence from
$\tauzerocut$ deteriorates substantially for  $|y| \gtrsim 2$.  However, even in the central region,
the $Z$ process is far more affected by  the choice of $\tauzerocut$ than the other two calculations.
In the more forward regions, which still contribute to the cross-section at an appreciable level, the
$\tauzerocut$ dependence rises to the level of a few tens of percent.  For this reason it is crucial to
apply the basic fiducial cuts introduced earlier in order to obtain a percent level agreement with the NNLO coefficient.

\begin{figure}[t]
\centering
\includegraphics[width=0.32\textwidth]{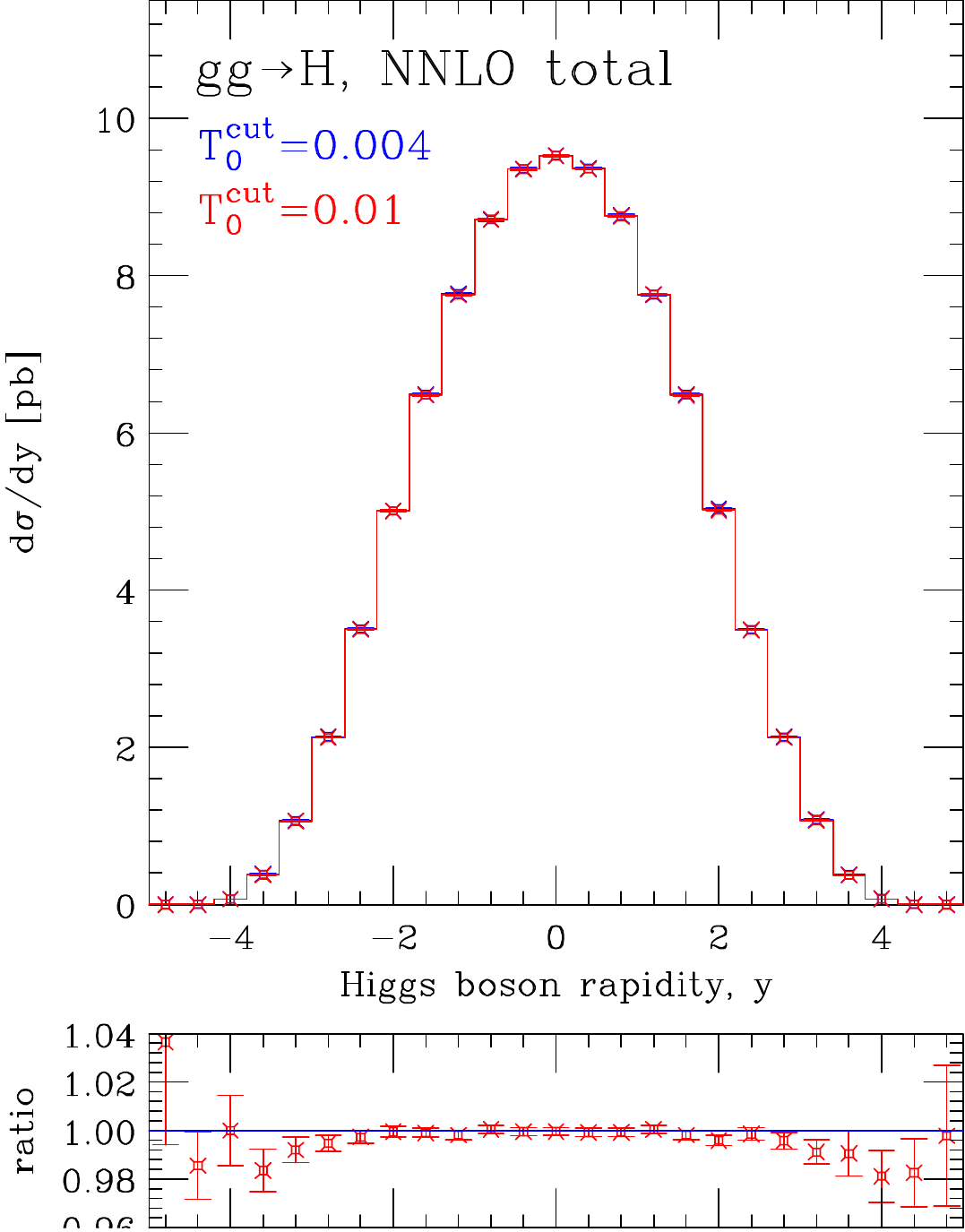}
\includegraphics[width=0.32\textwidth]{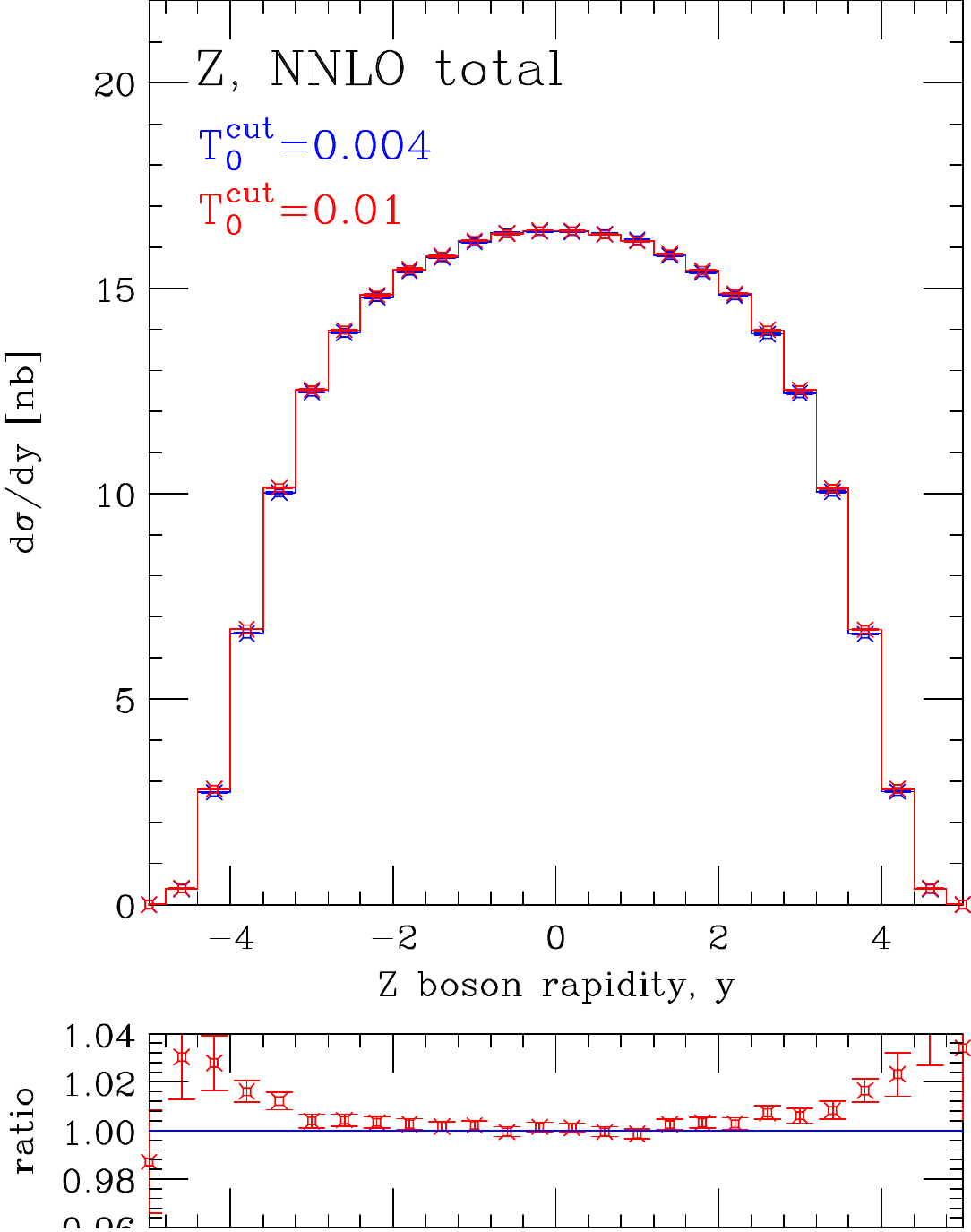}
\includegraphics[width=0.32\textwidth]{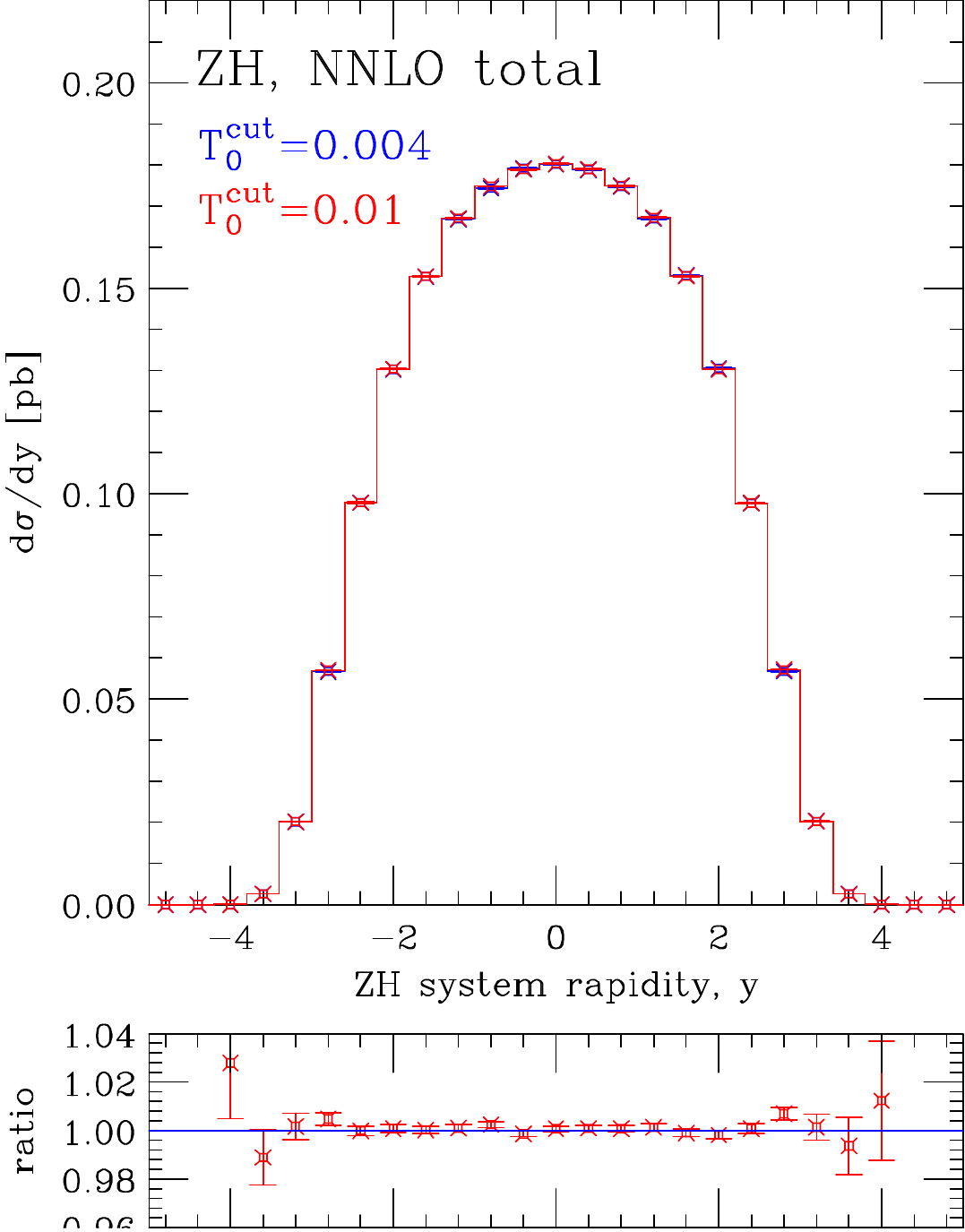}
\caption{\label{ynnlototal}MCFM calculation of the full NNLO result for the rapidity distributions
of the Higgs boson (left), $Z$ boson (centre) and $ZH$ system (right), in the $gg \to H$, $Z$ and $ZH$ production
processes respectively. Results are shown  for two values of $\tauzerocut$, with the lower panel showing the ratio of
the $\tauzerocut=0.01$ GeV result to the $\tauzerocut=0.004$ GeV one. }
\end{figure}

In contrast, for phenomenology it is sufficient to study the effect of the value of $\tauzerocut$ not on the
effect of the NNLO correction itself, but on the total prediction at that level of accuracy.  In that case
the smallness of the NNLO coefficient in the case of $Z$ production is an advantage as it suppresses the
relative size of the power corrections in the total.  On the other hand the $gg \to H$ process, which receives a
very large correction at NNLO, is more easily subject to power corrections.   In order to provide a full
NNLO prediction for the rapidity distributions discussed in this section we sum the results of a standard MCFM calculation
at NLO and a computation of only the NNLO correction using jettiness subtraction. The resulting distributions
are shown in Fig.~\ref{ynnlototal}.  The $gg \to H$ and $Z$ production processes differ by a couple of percent in the
tails of the distribution, for these two values of $\tauzerocut$, but are otherwise in excellent agreement.
The dependence on $\tauzerocut$ is even smaller for the case of $ZH$ production.

\subsection{Cross-sections under cuts}

Although the $W$ and $Z$ production cases are the most sensitive to
$\tauzerocut$ at NNLO, at this order both $ZH$ and $W^\pm H$ 
production also display a non-negligible dependence on 
$\tauzerocut$.  We therefore consider all four processes in this
section.  For $W$ and $Z$ production we apply the same cuts as before.
For the other processes we consider the final states
$W^\pm (\to e^\pm \nu) H(\to \gamma\gamma)$ and
$Z(\to e^+ e^-) H(\to b\bar b)$ but do not apply any cuts to the Higgs
boson decay products in either case.  In this way the results remain
valid for any decay channel of the Higgs boson.  The $W^\pm$ and $Z$
decay products are subject to the same cuts as in the corresponding inclusive
$W$ and $Z$ production processes.

\begin{figure}[t]
\centering
\includegraphics[width=11.cm]{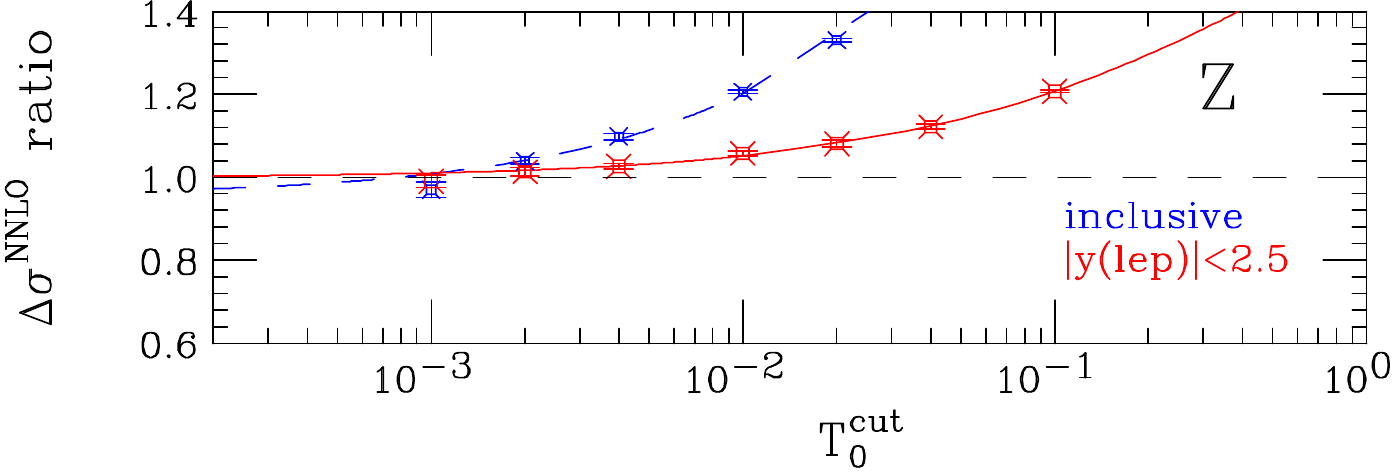} \\
\includegraphics[width=11.cm]{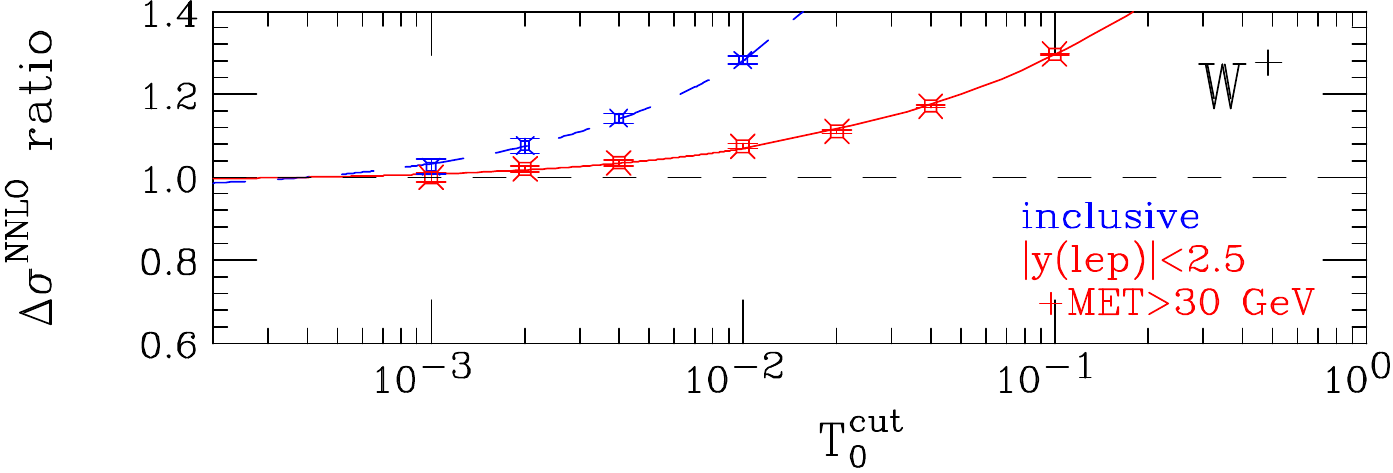} \\
\includegraphics[width=11.cm]{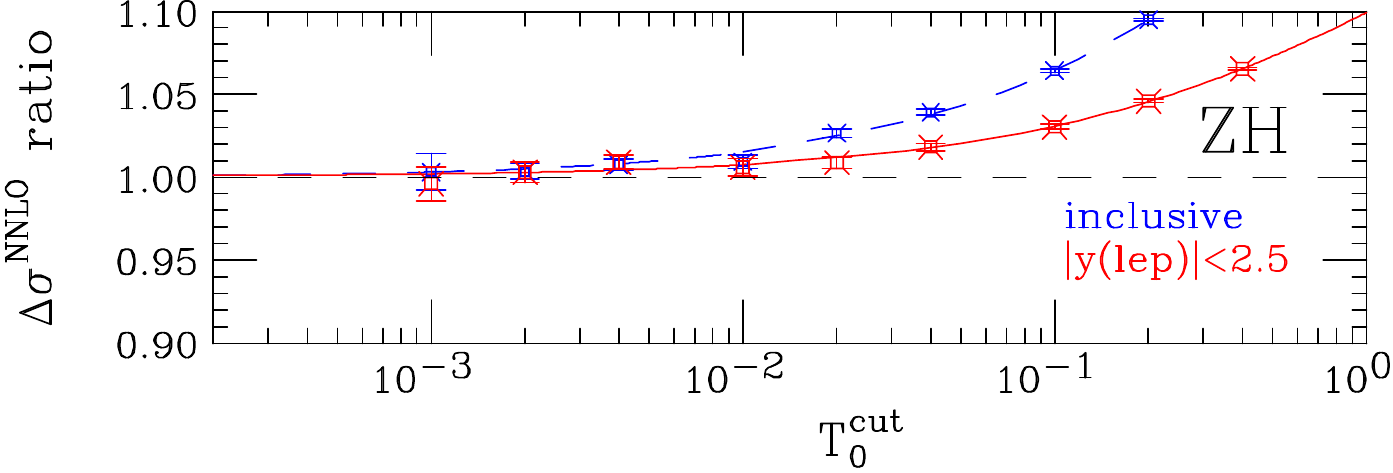} \\
\includegraphics[width=11.cm]{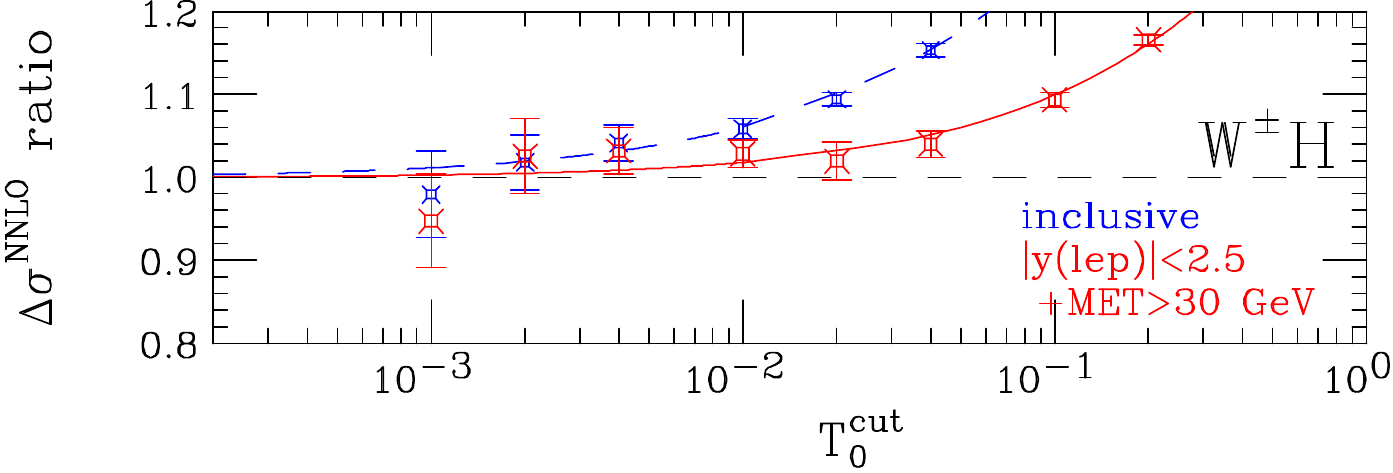} \\
\caption{\label{taudepnnlofid}The ratio of the MCFM $N$-jettiness calculation of the NNLO coefficient
to the (fitted) asymptotic result, as a function of the $N$-jettiness
resolution parameter $\tauzerocut$ (in GeV).  The comparison is performed for
$Z$, $W^+$, $ZH$ and $W^\pm H$ production and for both the inclusive case and for a minimal
set of fiducial cuts (detailed in the text). The lines represent fits to the individual
points using the form given in Eq.~(\ref{eq:nnloform}).}
\end{figure}

The results of this study are shown in Fig.~\ref{taudepnnlofid}.  For the
$W$ and $Z$ cases, the improvement is dramatic;  for
$\tauzerocut=0.02$ GeV the difference from the asymptotic result
improves from approximately $35$\% in the inclusive case to $8$\% under cuts.
A similar level of improvement applies in the case of $W$ production.
For $ZH$ production the gain is less pronounced due to the fact that only
the $Z$ decay products are restricted in rapidity, which results in a less
stringent constraint on the combined $ZH$ system.  Nevertheless, the agreement
with the asymptotic result improves by about a factor of two relative to the
case of no cuts. 
The asymptotic value of each NNLO $N$-jettiness calculation, together with
the  LO cross-sections under the fiducial cuts used in this study,
are shown in Table~\ref{NNLOfiducialcorrections}.
\begin{table}[t]
\begin{center}
\begin{tabular}{|l|r|r|c|}
\hline
Process    & $\sigma^{LO, fid}$
 & $\Delta\sigma^{NNLO, fid}$ & $\Delta\sigma^{NNLO, fid}/\sigma^{LO, fid}$ \\
\hline
$Z$        & 708.6 pb  	   & 44.8 pb    & 0.063 \\
$W^+$      & 3.259 nb 	   & 270 pb     & 0.083 \\
$ZH$       & 9.606 fb 	   & 1.126 fb   & 0.12 \\
$W^\pm H$  & 0.1337 fb 	   & 0.00353 fb & 0.026 \\
\hline
\end{tabular}
\caption{\label{NNLOfiducialcorrections}
NNLO corrections under the basic fiducial cuts described in the text.}
\end{center}
\end{table}

\section{ Analytic Power Corrections at  NLO and NNLO}\label{power}
\begin{figure}[t]
\centering
\includegraphics[width=0.65\textwidth]{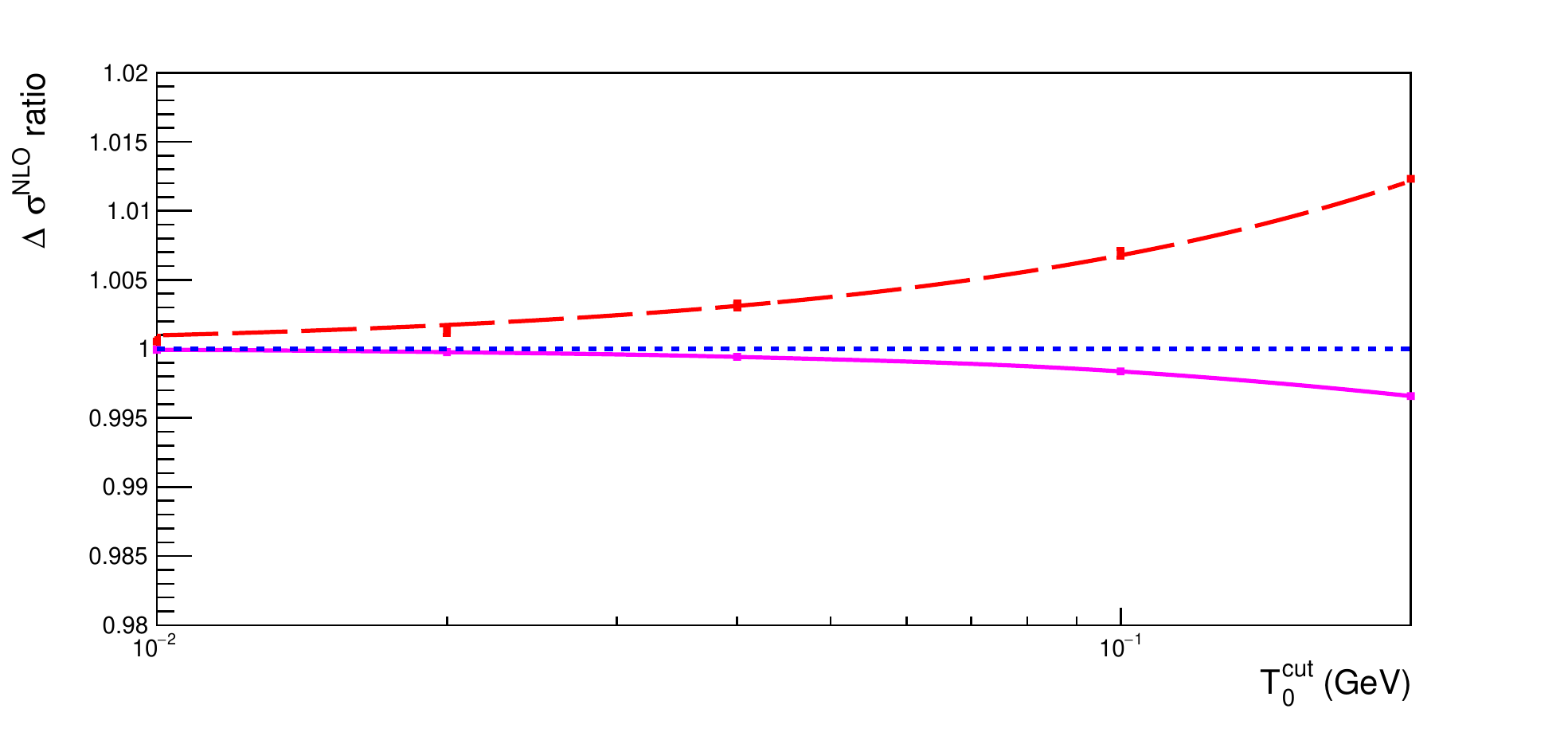}  
\includegraphics[width=0.65\textwidth]{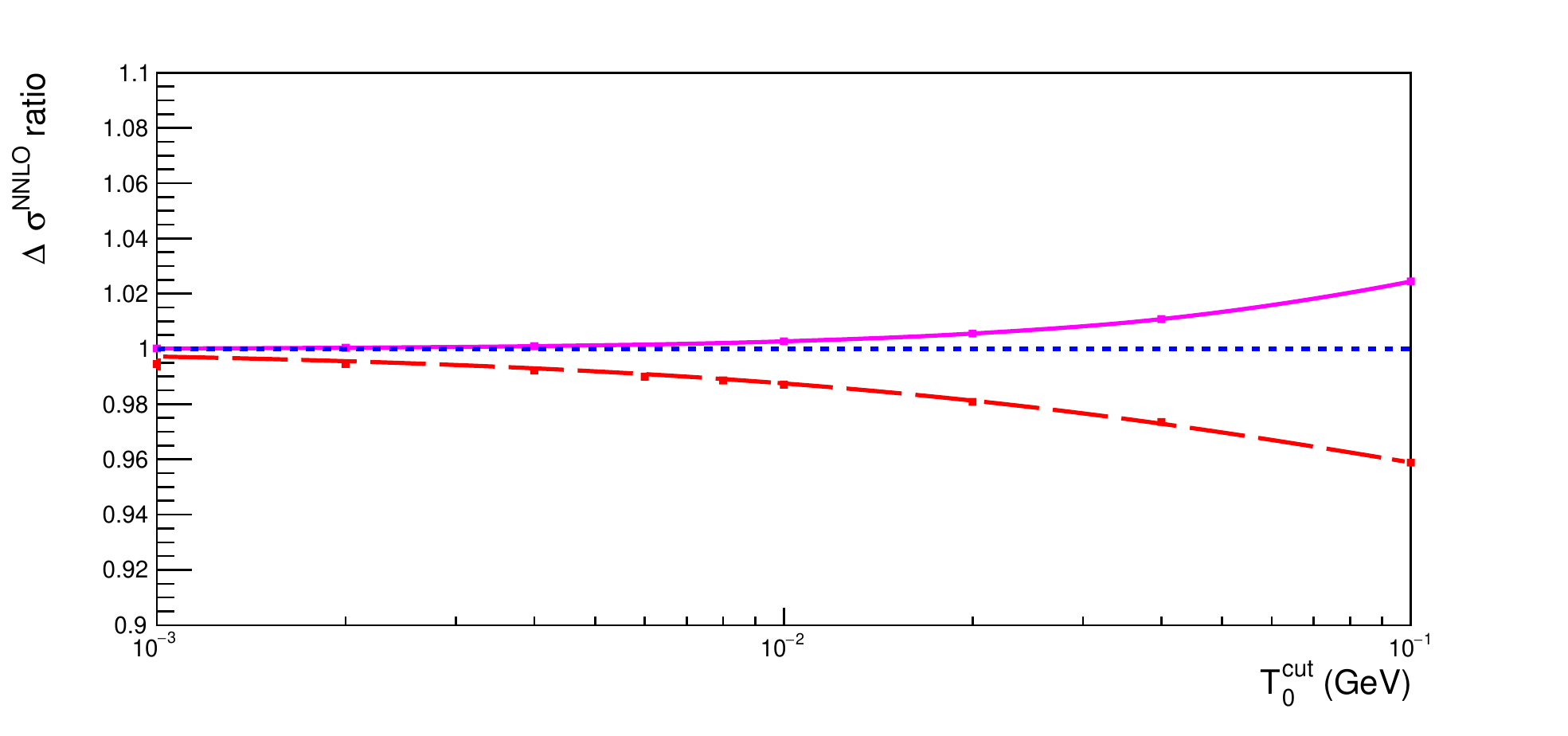}  \\
\caption{\label{powercor} 
The NLO $\tauzerocut$ dependence for $gg\rightarrow H$ of Fig.~\ref{taudepnlo}
is shown as the dashed red line in the upper pane. 
The solid purple line gives the
remaining $\tauzerocut$ dependence when the analytically calculated expression for the
leading power correction is added to MCFM.
The lower pane gives the same at NNLO for $gg\rightarrow H$ where the red line is 
taken from Fig.~\ref{taudepnnlo}.}
\end{figure}

As is clear from the discussions in the earlier sections it is important to understand the
$\tauzerocut$ dependence. The choice of this cut is a balance between achievable statistical
uncertainties and the uncertainty due to the missing power corrections of Eq.~(\ref{eq:nloform})
and Eq.~(\ref{eq:nnloform}). It would therefore be highly beneficial if one could calculate some of the
power corrections analytically. This will both speed up the code and increase the accuracy.

A preliminary calculation of the dominant power correction for NLO and NNLO 
as a function of the rapidity of the
Higgs boson has been performed~\cite{BPL}. The results for the total $gg\rightarrow H$ cross section
are shown in Fig.~\ref{powercor}. The full result for the leading power correction
to Drell-Yan like processes will be discussed in a separate publication~\cite{BPL} 
and included in a future version of MCFM.

As can be seen from Fig.~\ref{powercor} the effect of including this term
into MCFM is substantial and one can choose the $\tauzerocut$ an order of magnitude
larger and still obtain about the same uncertainty due to the new subleading power corrections.
This will have a large impact on the ultimate achievable precision of the jettiness
method as implemented in MCFM.

\section{Numerics}
\label{sec:comp}
In this section we discuss the numerical performance of MCFM. As an illustration
we will run the hybrid openMP/MPI version of MCFM 
on a modest sized cluster. This cluster consists of 24 nodes, each node having
of a motherboard with two Intel X5650 chips (2.67 GHz) using an unified memory. 
Each of the Intel chips has 6 cores, resulting in a total of $24\times
2\times 6=288$ computing cores for the cluster.
The nodes are connected using InfiniBand NFS mounts.

We will use $4\times 100,000+10\times 1,000,000$ VEGAS events in the remainder of this section.
It is straightforward to scale the results obtained for this particular cluster to other cluster configurations.
Specifically we examine two important performance issues. 
First, we will look at the time required to 
calculate the cross section as a function of the number of cores used. 
Second, we will look at the obtained statistical precision due to the Monte Carlo integration
as a function of the $\tauzerocut$ parameter.
For all the runs in this section we use, in addition to the input parameters of Table~\ref{param},
a collision energy of $14$ TeV and an inclusive anti-$k_T$ jet algorithm with a cone size of 0.4.
 We apply, where applicable, the following cut on the transverse momenta of the final state objects 
$p_T^{\mbox{\tiny JET}}>20$ GeV, $p_T^{l^\pm}>25$ GeV, $p_T^{\mbox{\tiny MISS}}>40$ GeV, 
$p_T^{\gamma_1}>40$ GeV and $p_T^{\gamma_2}>20$ GeV. The rapidity of all final state objects is
required to be less than 2.5 and we require a separation between the
observable final state objects of $\Delta R>0.4$.
When a $Z$-boson is produced we apply the additional cut on the di-lepton invariant mass of $40$ GeV 
with no separation requirement between the two charged leptons.

Calculating cross sections at higher order requires a significant amount of computing power. In Ref.~\cite{Campbell:2015qma} several of us extended MCFM to use openMP by modifying VEGAS in such a manner that it  distributes the event generation
and evaluation over the computing cores of a single node/motherboard. 
By using multiple computing cores openMP makes the evaluation of NLO 
cross sections on desktops efficient, while still using a single VEGAS grid for the optimization
of the numerical integration. 
For a timely evaluation of cross sections at NNLO it is desirable to use
a cluster combining many processors. As the processors in a cluster do not share the same physical memory
one has to use MPI. We extend VEGAS to use MPI to distribute the event generation
and evaluation further over all processors, while openMP still distributes the events per processor over its computing cores.
Again a single VEGAS grid is used to optimize the numerical integration.
It is important to use openMP to distribute the events on a single
processor as it keeps only one version of shared variables, while MPI
would keep  a separate copy of those variables for each MPI 
process thereby using the limited cache memory in an inefficient manner. This is particularly
important as MCFM use large shared arrays such as for example the VEGAS grid, PDF grids, histograms, etc.
which are common for all computing cores. It is therefore beneficial
to maintain a hybrid openMP/MPI version of VEGAS, especially given the continuing increase of the number of cores per processor.

There are two limits which come into play when executing parallel code.
The first limit is the memory bound limit. Here the evaluation time is determined by memory transfers
and not by computations. In this limit the evaluation time will not scale well when adding more computing cores
and improving the scaling behavior will be difficult, necessitating a better management of cache memory by the openMP code
and/or more efficient message passing by the MPI code. In the other limit the evaluation time is determined by the 
computations and time used for memory management is negligible. In this limit the execution time will scale 
perfectly with the number of processors, i.e. doubling the number of processors will half the execution time.
These limits are important in order to understand the scaling behavior seen in MCFM.
\begin{table}[t]
\begin{center}
\resizebox{\columnwidth}{!}{%
\begin{tabular}{|r||r|r||r|r||r|r|}
\hline
\multicolumn{1}{|c||}{} & \multicolumn{2}{c||}{LO}&\multicolumn{2}{c||}{NLO}&\multicolumn{2}{c|}{NNLO} \\
\cline{2-7}
cores & $W^+$ ($2\times 6$) & $W^+$ ($1\times 12$) & $W^+$ ($2\times 6$) & $W^+$ ($1\times 12$) & $W^+$ ($2\times 6$)&$W^+$ ($1\times 12$) \\ 
\hline
    6  & 11.15 &         & 308.9 &           & 10022 &         \\
  12  &   5.99 & 8.97 & 159.1 & 171.6 &   5068 & 5214 \\
  24  &   3.45 & 5.44 &   82.2 &   89.5 &   2559 & 2645 \\
  48  &   2.25 & 3.81 &   44.8 &   48.3 &   1326 & 1368 \\
  72 &    2.01 & 3.21 &   32.1 &   36.9 &     911 &   922 \\
  96 &    1.96 & 2.81 &   26.4 &   33.8 &     702 &   715 \\
144 &    1.95 & 2.47 &   23.7 &   29.4 &     505 &   506 \\
216 &    1.93 & 2.28 &   22.9 &   26.7 &     381 & 386   \\
288 &    2.20 & 2.32 &   23.3 &   25.4 &     328 & 358   \\
\hline
\end{tabular}
}
\caption{\label{Timing-W}
The time required (in seconds) to evaluate the $pp\rightarrow W^+\rightarrow l^+\nu$ total cross section
using the given number of cores for the node openMP texture of one MPI
job per processor ($2\times 6$) and a single MPI job per node ($1\times 12$). Results
are given for LO, NLO and NNLO total cross sections.}
\end{center}
\end{table}

\begin{figure}[t]
\centering
\includegraphics[width=13.cm]{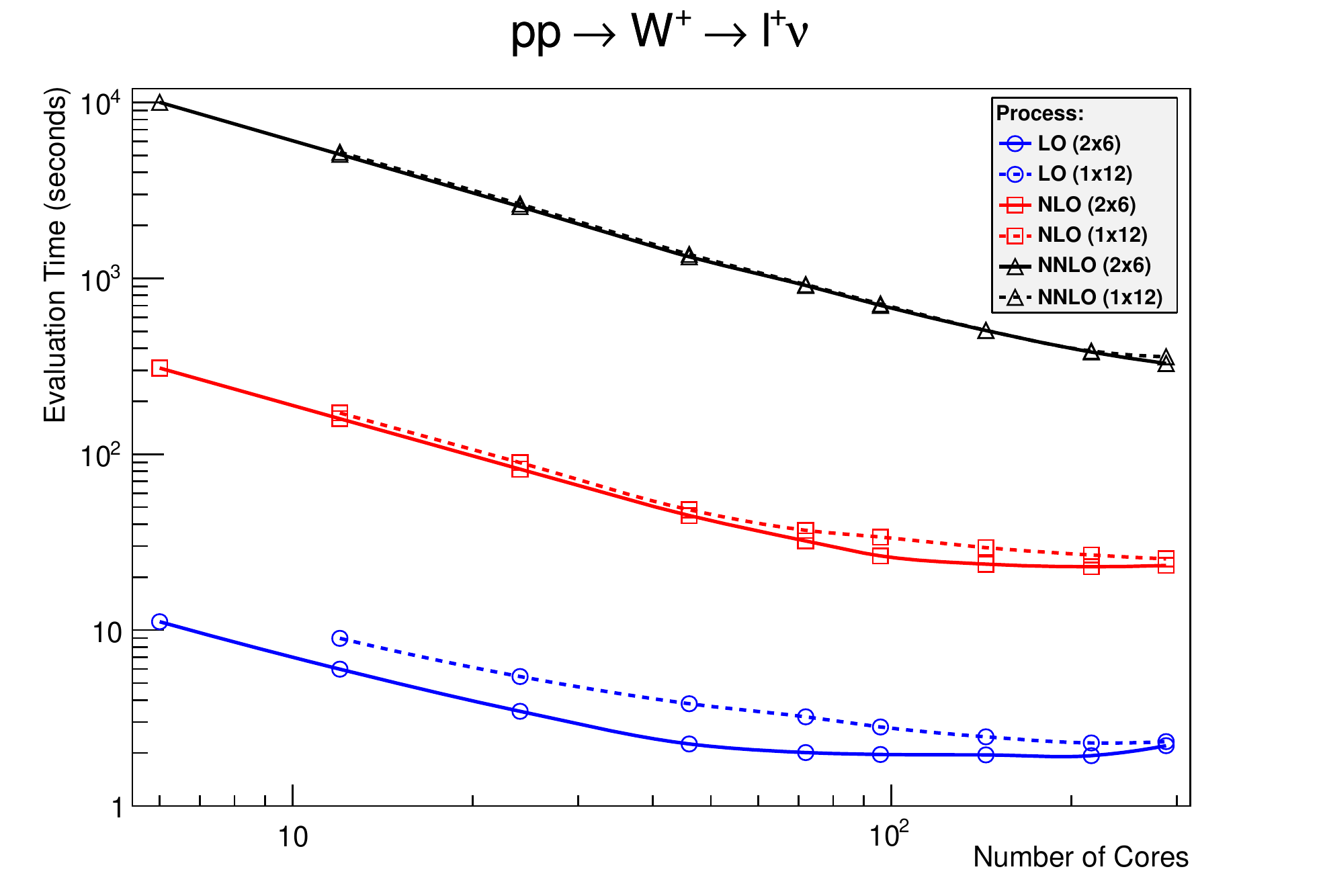} \\
\caption{\label{Scaling-W}
The evaluation time (in seconds) to calculate the total cross section
for the process $pp\rightarrow W^+\rightarrow l^+\nu$ at LO, NLO and NNLO
as a function of the number of cores used for both the $1\times 12$ and $2\times 6$ node texture.
}
\end{figure}
The scaling of the computing time with the number of processing cores for the process $pp\rightarrow W^+\rightarrow l^+\nu$
is given in Table~\ref{Timing-W}, with a visible representation in Figure~\ref{Scaling-W}.  
We have the option to run one MPI process per node and let openMP distribute the events over the 12 cores of
the two processors (indicated by the $1\times 12$ column). 
This in general is not a preferred mode of operating because the cache memory is divided over
the two processors, requiring openMP to make sure the two cache memories are synchronized, unduly invoking a
memory management overhead on the time needed for the evaluation. Alternatively, 
by running two MPI jobs per node openMP is used on a single processor
thereby optimizing the cache usage and minimizing memory management overhead ($2\times 6$ column).
This is clearly demonstrated in Table~\ref{Timing-W}. For example using all 288 processors on the cluster we can use of 48 MPI jobs 
(2 MPI jobs per node) to evaluate the NNLO cross section in 328 seconds, or use 24 MPI jobs (1 MPI job per node) requiring
358 seconds to evaluate the NNLO cross section.
The time difference is due to the fractured cache memory caused by forcing openMP to use two processors
in the case of running with 24 MPI jobs. 
Therefore in the remainder of this section we will use 2 MPI jobs per node, allowing openMP to operate on
a single processor.

Especially at LO, and to some extent at NLO, the computation effort to
evaluate this process is minimal, making the execution time operate
close to the memory bound limit.  This behavior is exhibited in
Table~\ref{Timing-W} and Figure~\ref{Scaling-W} where for LO the
evaluation time no longer improves when using more than 12 MPI jobs
(and for NLO more than 18 MPI jobs). At NNLO using more cores still
improves the evaluation time, as a consequence of the need to evaluate
a large number of the more computational intensive double parton
bremsstrahlung events.  It is worth noting that the execution time on
a single processor using openMP executes in under 3 hours making the
evaluation of this process on desktops very feasible. Running on the
full cluster using the 48 processors results in an execution time of
less than 4 minutes. This means one can easily increase the number of
events and lower the tau cut value to obtain better statistics.
\begin{table}[t]
\begin{center}
\begin{tabular}{|c||r|r|r|r|r|r|}
\hline
\multicolumn{1}{|c||}{MPI jobs} & \multicolumn{1}{c|}{$W^+$} & \multicolumn{1}{c|}{$Z$} & 
\multicolumn{1}{c|}{$H$} &  \multicolumn{1}{c|}{$HW^+$} & \multicolumn{1}{c|} {$HZ$}  
& \multicolumn{1}{c|}{$\gamma\gamma$} \\
\hline
 1  & 10022 & 20283 & 9079 & 9128 & 14357 & 27274\\
 2  &   5068 & 10173 & 4530 & 4639 &   7222 & 13704\\
 4  &   2559 &   5109 & 2339 & 2349 &   3655 &   6922\\
 8  &   1326 &   2581 & 1196 & 1211 &   1846 &   3492\\
12 &     911 &   1752 &   821 &   823 &   1276 &   2352 \\
16 &     702 &   1336 &   637 &   634 &     954 &   1773 \\
24 &     585 &     915 &   466 &   452 &     662 &   1219 \\
36 &     381 &     643 &   360 &   340 &     473 &     839\\
48 &     328 &     505 &   323 &   283 &     378 &     657\\
\hline
\end{tabular}
\caption{\label{Timing-all}
The time required (in seconds) to evaluate the 
$pp\rightarrow W^+\rightarrow l^+\nu$, $pp\rightarrow Z\rightarrow l^+l^-$,
$pp\rightarrow H\rightarrow \gamma\gamma$, $pp\rightarrow H+W^+\rightarrow\gamma\gamma+l^+\nu$,
$pp\rightarrow H+Z\rightarrow \gamma\gamma+l^+l^-$ and $pp\rightarrow\gamma\gamma$
cross sections at NNLO
using a given number of MPI processes for the node openMP texture of one MPI
job per 6-core processor ($2\times 6$).}
\end{center}
\end{table}

\begin{figure}[t]
\centering
\includegraphics[width=13.cm]{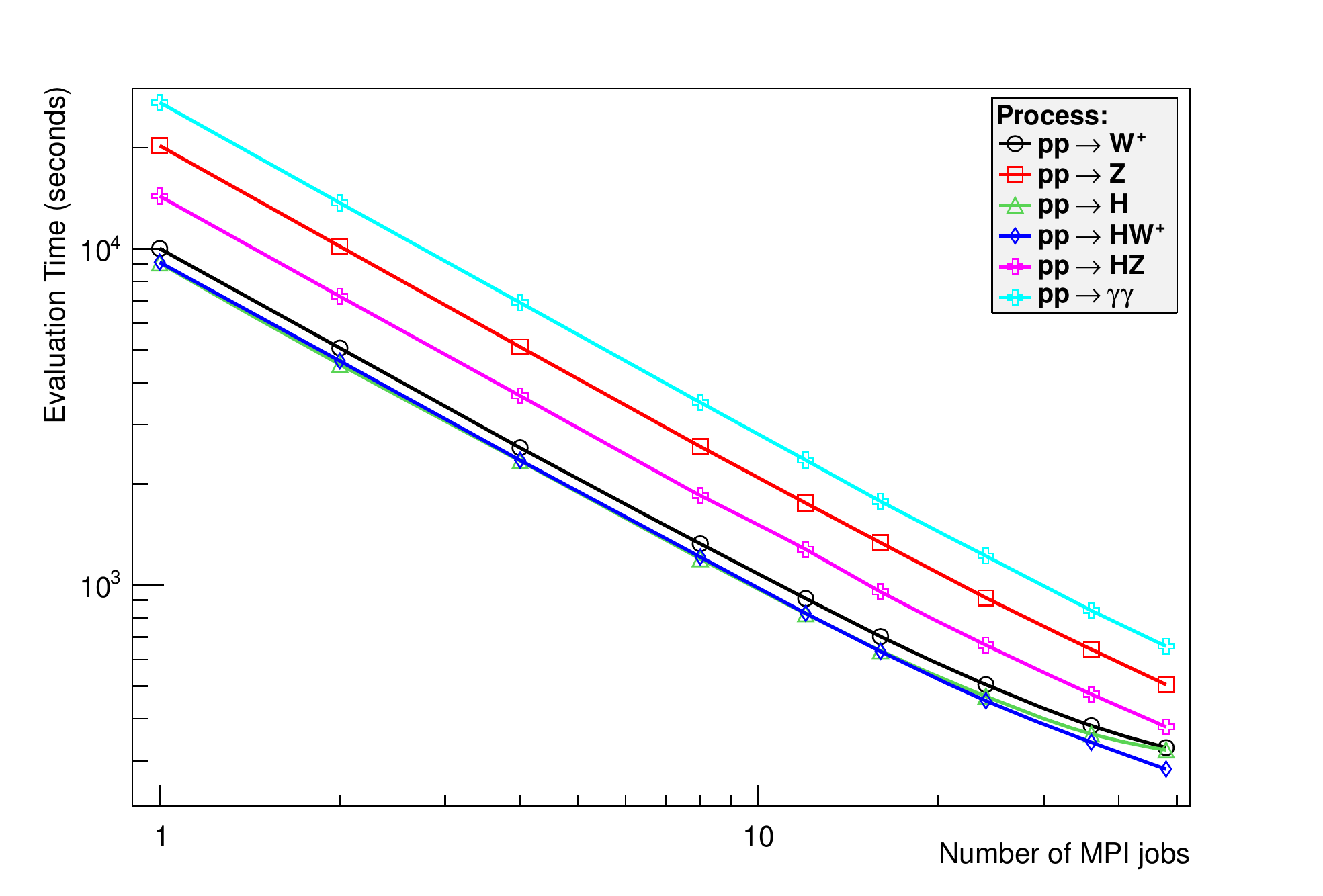} \\
\caption{\label{Scaling-all}
The evaluation time (in seconds) needed to calculate the total cross section
for the processes $pp\rightarrow W^+\rightarrow l^+\nu$,
$pp\rightarrow Z\rightarrow l^+l^-$, $pp\rightarrow H\rightarrow\gamma\gamma$,
$pp\rightarrow H+W^+\rightarrow\gamma\gamma+ l^+\nu$,
$pp\rightarrow H+Z\rightarrow\gamma\gamma+\l^+l^-$
and $pp\rightarrow\gamma\gamma$ at NNLO
as a function of the number of MPI processes used (each MPI process uses openMP on 6 cores).
}
\end{figure}
Next we look at all the new NNLO processes added to MCFM in Table~\ref{Timing-all}
and Figure~\ref{Scaling-all} where the time in seconds is given as a function of the
number of MPI jobs (= number of processors) used. As can be seen, the
processes scale well all the way up to the 288 processors. Some
indication of a less than perfect scaling can be seen in the simplest
of the NNLO processes $pp\rightarrow H\rightarrow\gamma\gamma$ when we
get to a high number of processors indicating there is some memory
overhead. All other processors still are computing dominated which
will allow easy speed-up by invoking even more processors. The most
complicated NNLO process $pp\rightarrow\gamma\gamma$ takes just under 11
minutes to evaluate using 48 processors. Therefore obtaining higher
statistics is rather easy. This process would still only take a bit less
than 8 hours on a single processor desktop.

\begin{table}[t]
\begin{center}
\resizebox{\columnwidth}{!}{%
\begin{tabular}{|r||r|r|r|r|r|r|}
\hline
\multicolumn{1}{|c||}{$\tauzerocut$} & \multicolumn{1}{c|}{$W^+$} & \multicolumn{1}{c|}{$Z$} & 
\multicolumn{1}{c|}{$H$} &  \multicolumn{1}{c|}{$HW^+$} & \multicolumn{1}{c|} {$HZ$}
& \multicolumn{1}{c|}{$\gamma\gamma$} \\
\hline
0.001  &      2\% (1397) &   0.9\% (2770) & 0.05\% (1256) &  10\% (1263)  &   6\% (1939)  & 0.4\% (3706) \\
0.005  &   0.7\% (1358)  &   0.4\% (2701) & 0.04\% (1234) &    3\% (1238) &   2\% (1906)  & 0.2\% (3661) \\
0.01   &   0.5\% (1356)  &   0.2\% (2677) & 0.04\% (1214) &    2\% (1222) &   1\% (1847)  & 0.15\% (3585) \\
0.05   &   0.2\% (1315)  & 0.08\% (2572)  & 0.04\% (1197) & 0.6\% (1206)  & 0.4\% (1841)  & 0.09\% (3492) \\
0.1    & 0.09\% (1307)   & 0.05\% (2526)  & 0.04\% (1186) & 0.3\% (1186)  & 0.2\% (1847)  & 0.08\% (3427) \\
0.5    & 0.04\% (1266)   & 0.04\% (2356)  & 0.04\% (1176) & 0.1\% (1150)  & 0.09\% (1768) & 0.07\% (3376) \\
\hline
\end{tabular}
}
\caption{\label{Precision}
The relative statistical precision (in percentages) on the $pp\rightarrow W^+\rightarrow l^+\nu$, $pp\rightarrow Z\rightarrow l^+l^-$,
$pp\rightarrow H\rightarrow \gamma\gamma$, $pp\rightarrow H+W^+\rightarrow\gamma\gamma+l^+\nu$
$pp\rightarrow H+Z\rightarrow \gamma\gamma+l^+l^-$ and $pp\rightarrow\gamma\gamma$
cross sections at NNLO as a function of $\tauzerocut$ (in GeV) using $4\times 2\times 6$ cores. 
Also given in brackets is the evaluation time (in seconds).
}
\end{center}
\end{table}
Finally, the statistical integration error obtained for the inclusive
cross section given the cuts using the 10,000,000 VEGAS events as a function
of the $\tauzerocut$ is given in
Table 10. The evaluation times are given for using 8 processors. As
can be seen there is a small dependence of the evaluation time on the
choice of the $\tauzerocut$. As we choose the $\tauzero$ cut smaller the Monte
Carlo becomes more ``efficient'' because it will generate more
soft/collinear events. That is, less events will be rejected by the
cuts hence the evaluation time will grow.

As can be seen from table~\ref{Precision}, the acquired statistical uncertainty is
quite process dependent. However the value of the $\tauzerocut$ will also
determine the systematic error due to the power corrections. Looking
at Table~\ref{summary} we see the required value of $\tauzerocut$ to reduce 
the power corrections to a 1\% or a 0.2\% level.\footnote{The size of the power corrections for
the diphoton process is obtained from the results of ref.~\cite{Campbell:2016yrh}.}
First, focussing on the 1\%
uncertainty we see that in all cases statistical error obtained with
the 10,000,000 events is smaller than 1\%. The worst case is the
inclusive $W^+$ production with a statistical uncertainty of
0.7\%. For all other cases the statistical error is more than on order
of magnitude smaller.
To achieve a systematic error of about 0.2\% we see that we need to
reduce the statistical uncertainty significantly in order to be
smaller than the systematic error. The reduction for some processes is
about an order of magnitude, requiring of the order of 100 times more
events. This means that an overall uncertainty of order $1\%$ is easily obtainable
using a desktop, however going to the per-mille level will require a
modest computer cluster such as the one used for the numerical results
in this section.
\begin{table}[t]
\begin{center}
\begin{tabular}{|rl|cc|}
\hline
Process &
 & \multicolumn{2}{|c|}{$\sigma^{NNLO}$} \\
& &  $1\%$ accuracy &  $0.2\%$ accuracy  \\
\hline
$gg \to H$ & inclusive  & 0.03 & 0.002 \\
\hline
$Z$ & inclusive  & 0.01 & 0.002 \\
    & lep. cuts  & 0.07 & 0.005 \\
\hline
$W^+$ & inclusive  & 0.005 & 0.001 \\
      & lep. cuts  & 0.03  & 0.003 \\
\hline
$ZH$ & inclusive  & 0.3 & 0.02 \\
     & lep. cuts  & 0.8 & 0.04 \\
\hline
$W^\pm H$ & inclusive & 0.2 & 0.01 \\
          & lep. cuts & 0.8 & 0.08 \\
\hline
$\gamma\gamma$ & cuts~\cite{Campbell:2016yrh} & 0.01 & 0.001 \\
\hline
\end{tabular}
\caption{\label{summary}
Values of $\tauzerocut$ (in GeV) required to perform 
the NNLO $N$-jettiness calculation to a given accuracy,
for the processes studied in this paper.
At larger $\tauzerocut$ the accuracy deteriorates
because of increased power corrections.}
\end{center}
\end{table}
%
%

\section{Summary}
\label{sec:conc}

In this paper we detailed the performance of the first NNLO version of
MCFM. Using the non-local $N$-jettiness subtraction method, we included the NNLO
corrections for six final states: $pp\rightarrow
W^\pm$, $pp\rightarrow Z/\gamma^*$, $pp\rightarrow H$, $pp\rightarrow W^\pm H$,
$pp\rightarrow ZH$ and $pp\rightarrow\gamma\gamma$. For each process decays of the unstable 
vector bosons are included where appropriate. The method was
checked at NLO against existing calculations and excellent agreement
was found. At NNLO the dependence on the jettiness cut
was studied in great detail and some guidelines on the choice of the
jettiness cut were given for all NNLO processes added to MCFM.

Another addition to MCFM is the ability to run in a hybrid openMP/MPI
mode, enabling the Monte Carlo to use clusters efficiently while still
maintaining a single VEGAS grid. The evaluation time for NNLO
inclusive cross sections with an overall precision better than 1\% on
a single 8-core processor using openMP ranges from 3 hours to 8 hours
depending on the specific process. It was shown all processes scaled
well on a multi-processor cluster using MPI in addition to openMP,
giving an evaluation time of 5 minutes to 11 minutes on a 48 8-core
processor cluster.  A small cluster of 50+ cores will give good
statistics for distributions at NNLO in a short time-frame.

The $N$-jettiness subtraction method in MCFM is now
well tested. Combined with the hybrid openMP/MPI option, MCFM
is now ready to include in the near future more complicated processes
including jet final states.
  
\acknowledgments
RKE would like to thank Robert Harlander for helpful discussions.  R.~B. is supported by the DOE contract DE-AC02-06CH11357.  J.~C. and W.~G. are supported by the DOE contract DE-AC02-07CH11359.  C.~F. is supported by the NSF grant PHY-1520916.  X.~L. is supported by the DOE grant DE-FG02-93ER-40762.  F.~P. is supported by the DOE grants DE-FG02-91ER40684 and DE-AC02-06CH11357. Support was provided by the Center for Computational Research at the University at Buffalo.   R.~B., W.~G., X.~L., and F.~P. thank the Kavli Institute for Theoretical Physics at the University of California, Santa Barbara, which is supported by the U. S. National Science Foundation under Grant No. NSF PHY11-25915.

\bibliographystyle{JHEP}
\bibliography{Singlet}
\end{document}